\documentclass[journal]{IEEEtran}
\ifCLASSINFOpdf
\else
\fi
\usepackage{gensymb}
\usepackage{amsfonts}
\usepackage{amsmath}
\usepackage{multirow}
\usepackage{booktabs}
\usepackage{subfigure}
\usepackage{enumitem}
\usepackage{graphicx}
\usepackage{caption}
\usepackage{threeparttable}
\begin{document}
%
\title{Adaptive Hypergraph Convolutional Network for No-Reference 360-degree Image Quality Assessment}
%
%
%

\author{Jun Fu,
        Chen Hou,
        Wei Zhou,
        Jiahua Xu,
        and~Zhibo Chen,~\IEEEmembership{Senior Member,~IEEE}
\thanks{J. Fu, C. Hou, W. Zhou,  J. Xu and Z. Chen are with the CAS Key Laboratory of Technology in GeoSpatial Information Processing and Application System, University of Science and Technology of China, Hefei 230027, China (e-mail: fujun@mail.ustc.edu.cn;divide@mail.dlut.edu.cn; weichou@mail.ustc.edu.cn;xujiahua@mail.ustc.edu.cn chenzhibo@ustc.edu.cn).}}

%
%

\markboth{Submission to IEEE Transactions on Circuits and Systems for Video Technology}%
{Shell \MakeLowercase{\textit{et al.}}: Bare Demo of IEEEtran.cls for IEEE Journals}
%



\maketitle

\begin{abstract}
In no-reference 360-degree image quality assessment (NR 360IQA), graph convolutional networks (GCNs), which model interactions between viewports through graphs, have achieved impressive performance. However, prevailing GCN-based NR 360IQA methods suffer from three main limitations. First, they only use high-level features of the distorted image to regress the quality score, while the human visual system (HVS) scores the image based on hierarchical features. Second, they simplify complex high-order interactions between viewports in a pairwise fashion through graphs. Third, in the graph construction, they only consider spatial locations of viewports, ignoring its content characteristics. Accordingly, to address these issues, we propose an adaptive hypergraph convolutional network for NR 360IQA, denoted as AHGCN. Specifically, we first design a multi-level viewport descriptor for extracting hierarchical representations from viewports. Then, we model interactions between viewports through hypergraphs, where each hyperedge connects two or more viewports. In the hypergraph construction, we build a location-based hyperedge and a content-based hyperedge for each viewport. Experimental results on two public 360IQA databases demonstrate that our proposed approach has a clear advantage over state-of-the-art full-reference and no-reference IQA models.
\end{abstract}


\section{Introduction}
360-degree images/videos, also known as omnidirectional, panoramic, or VR images/videos, have become increasingly popular with the boost of virtual reality (VR) technology~\cite{fu2020sequential}. Compared to 2D images/videos, 360-degree ones allow users to interactively manipulate the perspective through Head Mounted Devices (HMDs), thereby bringing users an immersive experience. In real VR broadcasting systems, 360-degree images/videos typically undergo three stages, i.e., acquisition, compression, and transmission, before reaching end users. Among these stages, 360-degree images/videos may be degraded by various distortions such as white noise, blurring, and compression artifacts~\cite{gu2014hybrid,sun2021graphiqa}. The quality degradation of 360-degree content may significantly impair the user's quality of experience (QoE)~\cite{zhang2018subjective}. Therefore, it is important to study the quality assessment of 360-degree images, which can be used as guidance for optimizing existing VR broadcasting systems.

Compared with its 2D counterpart, 360-degree image quality assessment (360IQA) encounters more challenges. First, 360IQA needs to consider geometric deformation and pixel redundancy issues introduced in storing 360-degree contents. Second, 360IQA needs to consider the gap between the viewport image seen in the HMD and its corresponding 360-degree image in the equirectangular format. Third, 360IQA needs to consider interactions between viewports. Specifically, since they only watch a small portion of the 360-degree image at a time, users need to browse multiple viewports for accurate quality assessment. During this viewing process, users assess each viewport based on the visual information of its neighboring viewports and then obtain the final quality score through viewport quality aggregation.
\begin{figure}[tbp]
	\begin{tabular}{ccc}
		\includegraphics[width=0.3\linewidth]{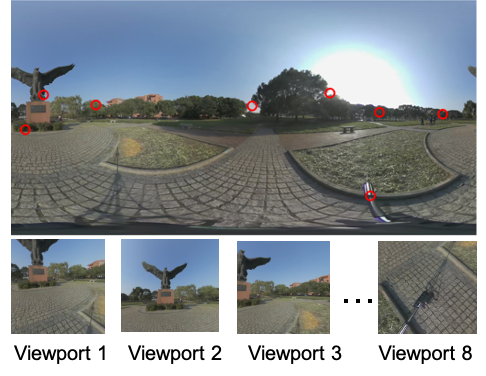} & \includegraphics[width=0.3\linewidth]{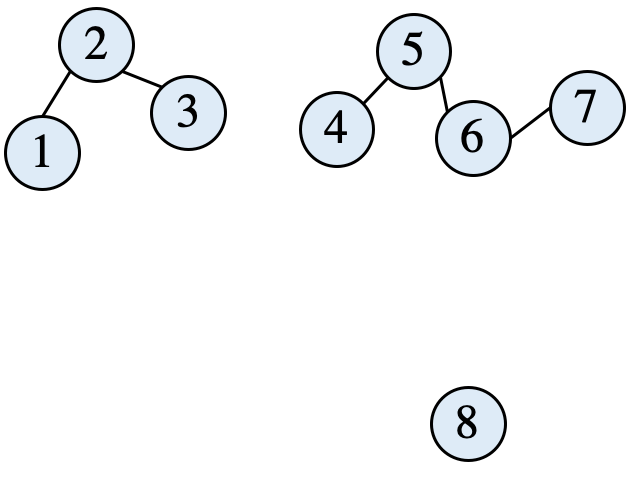} & \includegraphics[width=0.3\linewidth]{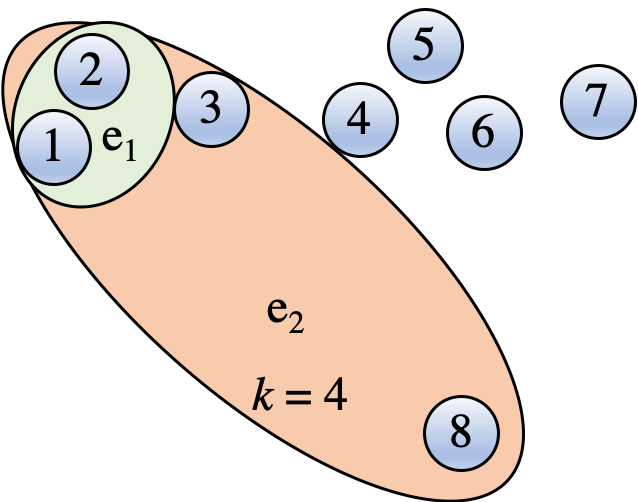} \\
		(a) & (b) & (c) 
	\end{tabular}
	\caption{Comparison of graphs and hypergraphs in modeling interactions between viewports.
		(a) The distorted 360-degree image and 8 key viewports.  (b) The constructed graph based on spatial locations of viewports. (c) Two hyperedges constructed for the node ``1''. ``e$_1$'' and ``e$_2$'' contain spatial and semantic neighborhoods of the node ``1'', respectively. ``$k$'' denotes the number of semantic neighborhoods.}
	\label{fig:Comparison}
\end{figure}

For this reason, in the past few years, lots of efforts have been made to overcome the challenges in 360IQA. Typically, 360IQA can be divided into two categories: full-reference (FR) 360IQA and no-reference (NR) 360IQA. For FR 360IQA, due to geometry deformation and pixel redundancy problems, it is infeasible to directly use FR 2DIQA metrics such as PSNR and SSIM~\cite{ssim}. Accordingly, researchers design various variants of FR 2DIQA metrics, e.g., WS-PSNR~\cite{WSPSNR} and WS-SSIM~\cite{WSSSIM}. For NR 360IQA, early studies use patches cropped from the distorted 360-degree image to predict the quality score. As patches are not the actual content viewed by users, viewport-oriented methods are thus developed. Recently, the viewport-based graph convolutional neural network (VGCN~\cite{xu2020blind}) considers interactions between viewports and model them using graphs. In VGCN, the graph structure is defined based on spatial locations of viewports (as shown in Fig.~\ref{fig:Comparison}(b)), and node features correspond to high-level features of viewports.

However, there are three main limitations in existing GCN-based NR 360IQA methods. First, only high-level features of viewports are used for quality evaluation, which is inconsistent with the human visual perception process. Specifically, it is known that the human brain hierarchically processes the perceived image and the HVS comprehends the image based on the obtained hierarchical features~\cite{van1983hierarchical}. As intrinsically related to image understanding, quality evaluation also relies on hierarchical features. Second, since each edge only connects two viewports, graphs have limited capabilities for modeling complicated interactions between three or more viewports. Third, the graph structure only represents spatial relations between viewports. However, it is also important for quality evaluation to consider semantic correlations between viewports. For example, as shown in Fig.~\ref{fig:Comparison}(b), although far away from the node ``1'', the node ``8'' can offer guidance to assess the quality of the road in the node ``1''. 

To address these issues, in this paper, we propose an adaptive hypergraph convolutional network for NR 360IQA, denoted as AHGCN. Specifically, we first develop a multi-level viewport descriptor, which combines low-level, mid-level, and high-level features of viewports to produce hierarchical representations. Then, we model interactions between viewports through hypergraphs instead of graphs. For each viewport, we construct a location-based hyperedge based on the angular distance between viewports, and a content-based hyperedge according to the content similarity between viewports.  Experimental results on two public 360IQA databases demonstrate that the proposed AHGCN predicts the perceptual quality more accurately under various distortion types and levels compared with state-of-the-art 360IQA models. The main contributions of the proposed method are listed as follows:
\begin{itemize}
	\item We design a multi-level viewport descriptor and verify the effectiveness of hierarchical representations for 360IQA.
	\item We present the first attempt to employ hypergraphs for modeling interactions between viewports and show that hypergraphs are superior to graphs in capturing high-order dependencies.
	\item We propose an adaptive hyperedge construction method, which considers both spatial locations and content characteristics of viewports.
\end{itemize}
The rest of this paper is organized as follows. Section II introduces works related to our approach. We detail the proposed AHGCN for NR 360IQA in Section III, followed by experimental results presented in Section IV. Section V concludes the paper and points out some future directions. 
\begin{figure*}[htbp]
	\includegraphics[width=\linewidth]{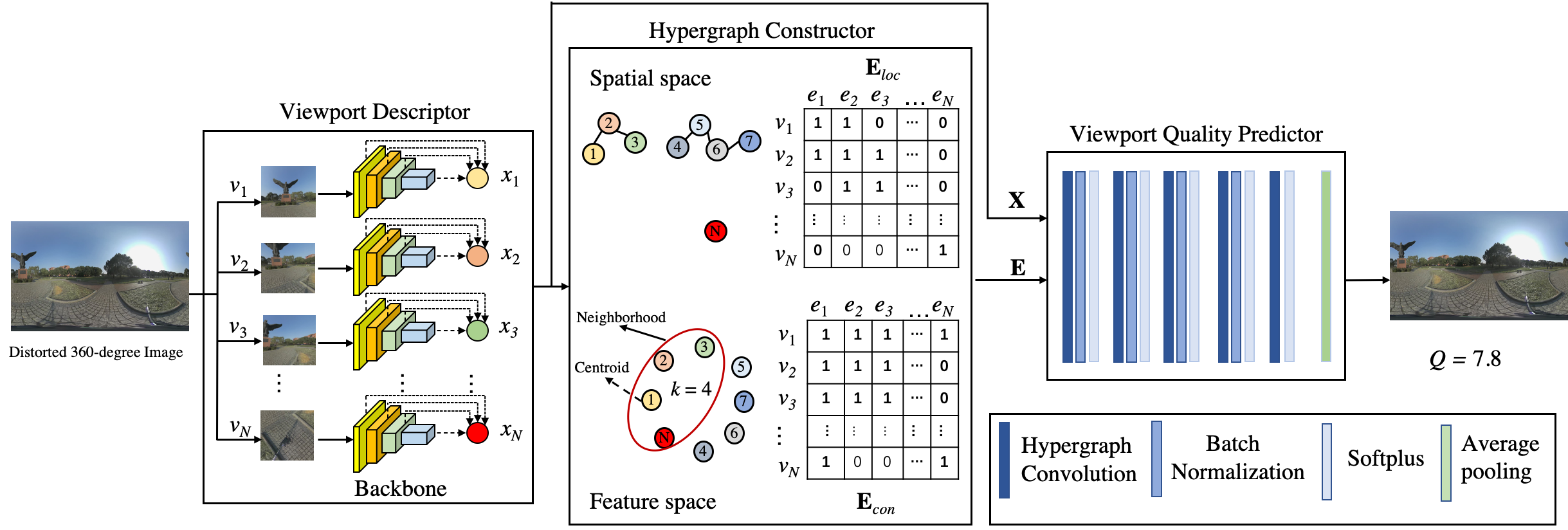}	
	\caption{Illustration of the proposed adaptive hypergraph convolutional network. \textbf{E} and \textbf{X} denote the hypergraph structure and hierarchical features of viewports. \textbf{E} consists two parts: location-based hyperedges \textbf{E}$_{loc}$ and content-based hyperedges \textbf{E}$_{con}$. $Q$ is the predicted quality score. ``$k$'' represents the number of semantic neighborhoods.}
	\label{fig:framework}
\end{figure*}
\section{Related Work}
In this section, we first overview FR 360IQA metrics, then introduce NR 360IQA models, and finally present the progress of the hypergraph learning.
\subsection{FR 360IQA}
FR 360IQA aims to evaluate the quality of the distorted 360-degree image given the original version. Compared to FR 2DIQA, FR 360IQA needs to consider geometric deformation and pixel redundancy issues brought in storing 360-degree content. To address this issue, diverse variants of FR 2DIQA models are proposed. Yu et al.~\cite{SPSNR} develop a spherical PSNR (S-PSNR), which calculates PSNR for the set of points uniformly distributed on a sphere instead of the rectangular plane. Sun et al.~\cite{WSPSNR} design a weighted spherical PSNR (WS-PSNR), which weights the error of points sampled on the 2D plane according to its stretch degree. Zakharchenko et al.~\cite{CPPPSNR} propose the Craster Parabolic Projection PSNR (CPP-PSNR), which calculates PSNR on the Craster Parabolic Projection domain. Xu et al.~\cite{CPPCNR} put forward a non-content-based PSNR (NCP-PSNR), which weights different regions based on the distribution of viewing directions. Like PSNR, SSIM~\cite{ssim}  also derives various versions for FR 360IQA. Chen et al.~\cite{SSSIM}  propose a spherical SSIM (S-SSIM), which computes the similarity between distorted and original 360-degree images on the sphere. Zhou et al.~\cite{WSSSIM} develop weighted-to-spherically-uniform SSIM (WS-SSIM). Facebook~\cite{ssim360} designs SSIM360, which overcomes the warping issue of 360-degree images by a weighting mechanism.
\subsection{NR 360IQA}
NR 360IQA aims to score the distorted 360-degree image without the reference image. Early NR 360IQA methods usually leverage patch-level features to perform quality prediction. Kim et al.~\cite{DeepVRIQA1,DeepVRIQA2} propose a  deep-learning-based VR image quality assessment framework (DeepVR-IQA) with adversarial learning. It firstly predicts quality scores of patches sampled from the distorted 360-degree image and then fuses them to obtain the final quality score using a position-aware weighting mechanism. Similarly,  Li et al.~\cite{VQA-ODV} predict quality scores of patches and weight them based on the estimated head movement (HM) and eye movement (EM) maps. Considering the inconsistency between the viewport image and the patch, viewport-based NR 360IQA methods are thus proposed. Li et al.~\cite{VCNN} design a multi-task framework, which simultaneously predicts quality scores of viewports and performs HM and EM map prediction. Zhou et al.~\cite{zhou2021no} present a NR 360IQA framework, which predicts the quality score of the distorted 360-degree image based on multi-frequency information and local-global naturalness. Sun et al.~\cite{mc360iqa2} put forward a multi-channel convolution neural network for blind 360-degree image quality assessment (MC360IQA). It uses a shared ResNet-34 network to parallelly extract high-level features from six viewports, and concatenate them for quality score regression. Recently, considering the importance of interactions between viewports, Xu et al.~\cite{xu2020blind} develop a viewport-oriented graph convolutional network (VGCN). Despite achieving remarkable performance, GCN-based NR 360IQA methods still suffer from several limitations as aforementioned. 

\subsection{Hypergraph Learning}
Unlike conventional graphs, where each edge only connects two nodes, hypergraphs allow edges to connect three or more nodes. As such, hypergraphs are widely used to model complicated systems with high-order interactions. Zhou et al.~\cite{zhou2006learning} firstly introduce hypergraphs into clustering, embedding, and classification tasks, and achieve superior performance than graph-based methods. Feng et al.~\cite{feng2019hypergraph} propose a hypergraph neural network (HGNN), which extends hypergraph-based representation learning to learn-based models. Jiang et al.~\cite{jiang2019dynamic} develop a dynamic hypergraph neural network (DHGNN), which uses k-NN and k-means clustering methods to construct hyperedges without a pre-defined hypergraph structure. Zhang et al.~\cite{zhang2019hyper} put forward a self-attention-based graph neural network for hypergraphs (Hyper-SAGNN), which learns an aggregation function for each hyperedge. 

Most of the existing studies on hypergraphs focus on classification tasks but pay less attention to other tasks. In this paper, we are the first to introduce hypergraphs to NR 360IQA for modeling interactions between viewports. Moreover, we develop an adaptive hyperedge construction method, which considers both spatial locations and content features of viewports.

\section{Proposed Method}
In this section, we first define the problem of viewport-based NR 360IQA. Then, we introduce the proposed AHGCN in detail. Finally, we describe the implementation and training details.  
\subsection{Problem Formulation} 
Given $N$ viewports $\mathbf{V} = \{v_1,...,v_N\}$ sampled from a distorted 360-degree image $I$, we aim to predict the quality score of $I$ using a mapping function $F$:
\begin{equation}
Q = F(\mathbf{V} ;\theta),
\end{equation}
where $\theta$ represents all learnable parameters of the model $F$.

In this paper, we propose an adaptive hypergraph convolutional network (AHGCN) to model  $F$. As shown in Fig.~\ref{fig:framework},  AHGCN consists of three parts: a viewport descriptor, a hypergraph constructor, and a viewport quality predictor. Next, we will detail these parts in sequence.

\begin{figure}[htbp]
	\includegraphics[width=1\linewidth]{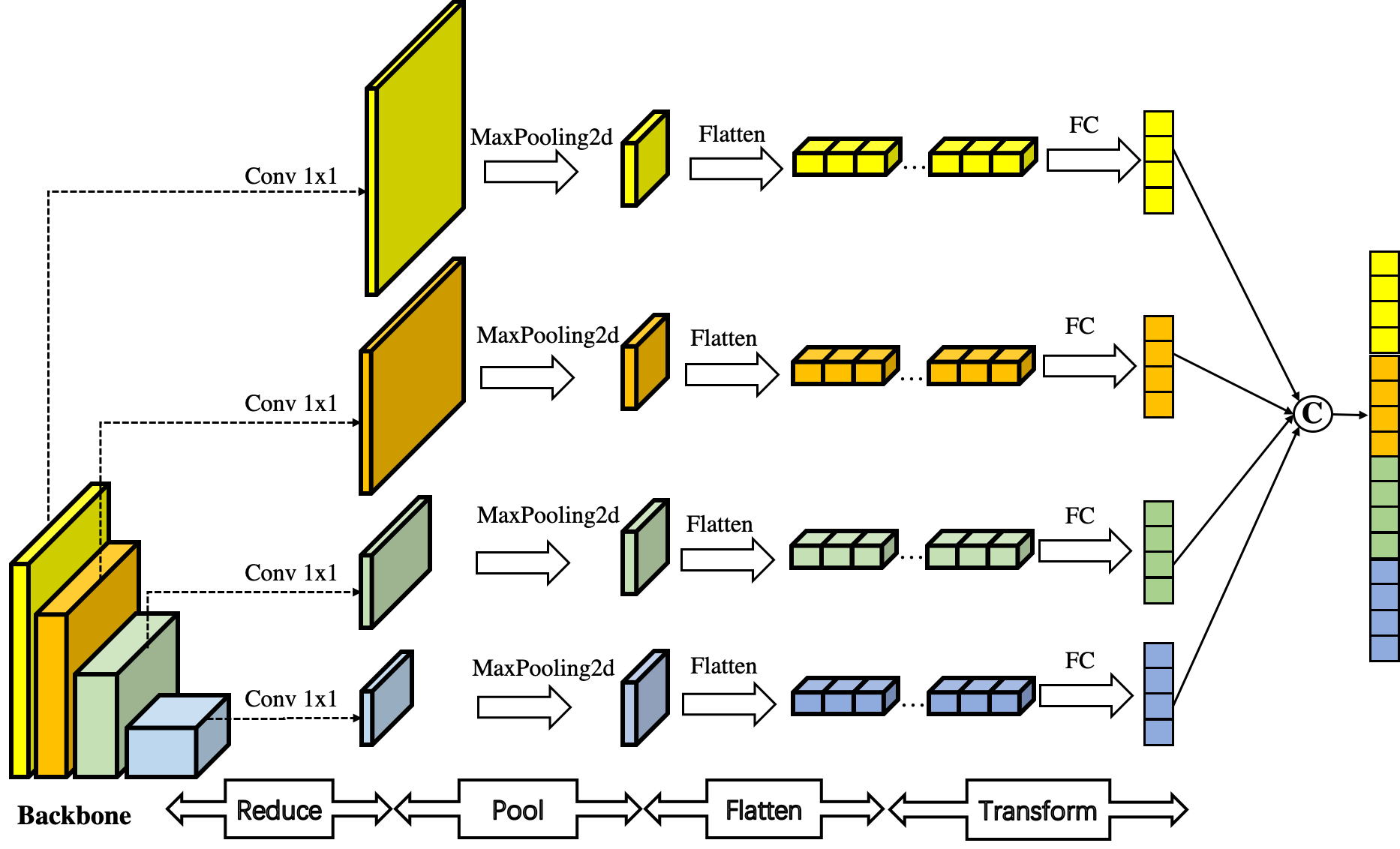}	
	\caption{The framework of the proposed viewport descriptor. ``Conv 1x1'' means the convolutional layer with the kernel size of 1, and ``FC'' denotes the fully connected layer.}
	\label{fig:viewport descriptor}
\end{figure}
\subsection{Viewport Descriptor}
The viewport descriptor aims to extract hierarchical features from the input viewport. Recently, deep convolutional neural networks have exhibited impressive power in representing perceptual image distortions~\cite{zhang2018unreasonable,prashnani2018pieapp,bosse2017deep}. Inspired by this, we bulid the viewport descriptor on backbones, pre-trained for object recognition~\cite{krizhevsky2012imagenet} on the ImageNet database~\cite{deng2009imagenet}. As shown in Fig.~\ref{fig:viewport descriptor}, the viewport descriptor first feeds the input viewport $v_i$ into a backbone network $f$:
\begin{equation}
\textbf{L}_i = \{l_{i,1},...,l_{i,m}\} = f(v_i;\theta_f), \forall i \in [1, N],
\end{equation}
where $ \textbf{L}_i$ denotes multi-level features of the $i$-th viewport, $l_{i,m}$ represents the feature map at the $m$-th level, and $\theta_f$ represents pre-trained weights of the backbone. Then, it individually compacts each feature map into a vector:
\begin{equation}
c_{i,j} = g(l_{i,j}),  \forall j \in [1, m],
\end{equation}
where the function $g(\cdot)$ denotes the principle of ``\textit{reduce-pool-flatten-transform}", $c_{i,j} \in \mathbb{R}^{d}$ is the compacted result of the feature map at $j$-th level. Finally, it concatenates compacted multi-level features to obtain hierarchical representations of the $i$-th viewport $x_{i} \in \mathbb{R}^{md}$:
\begin{equation}
x_{i} = c_{i,1} \cup c_{i,2} \cup ... \cup c_{i, m}.
\end{equation}

\subsection{Hypergraph Constructor}
To assess the distorted 360-degree image, subjects will browse the landscape of the sphere for a while. In this viewing process, the visual information of different viewports is interacted and aggregated for local quality evaluation~\cite{xu2020blind}. In this paper, we model interactions between viewports using hypergraphs, which can capture high-order interaction by connecting multiple viewports through one hyperedge. 

The hypergraph constructor aims to discover a location-based hyperedge and a content-based hyperedge for each viewport. The motivation of this design is two-fold. On the one hand, when evaluating one viewport, the user tends to refer to viewports close to it rather than those far away from it. This is mainly because dramatic head movements rarely occur in the user's viewing process. On the other hand, the visual information of viewports with same or similar content to the evaluated viewport may be also useful. For example, as shown in Fig.~\ref{fig:framework}, the visual information of the road in $N$-th viewport could offer guidance to assess the quality of the first viewport.
\\
\\
\noindent\textbf{Location-based Hyperedges}. The location-based hypergraph $\textbf{E}_{loc} = \{e^{1}_{loc},...,e^{N}_{loc}\} \in \mathbb{R}^{N\times N}$ consists of $N$ hyperedges, and the hyperedge $e^{i}_{loc} \in \mathbb{R}^{1\times N}$ collects spatial neighborhoods of the $i$-th viewport :
\begin{equation}
e^{i,p}_{loc} = \begin{cases}
1, AngularDist[(\phi_i, \Theta_i), (\phi_p, \Theta_p)] \leq \delta\\
0, otherwise
\end{cases},
\end{equation}
where the function $AugularDist(\cdot)$ computes the angular distance between the $i$-th viewport and the $p$-th viewport. $(\phi_i, \Theta_i)$ and $(\phi_p, \Theta_p)$ denote the longitudes and latitudes of viewport $i$ and $p$. $\delta$ is a pre-defined angular distance threshold. 
\\
\\
\noindent\textbf{Content-based Hyperedges}. The content-based hypergraph $\textbf{E}_{con} = \{e^{1}_{con},...,e^{N}_{con}\} \in \mathbb{R}^{N\times N}$ also consists of $N$ hyperedges, and the hyperedge $e^{i}_{con} \in \mathbb{R}^{1\times N}$ collects semantic neighborhoods of the $i$-th viewport :
\begin{equation}
e^{i,p}_{con} = \begin{cases}
1, v_p \in \mathcal{N}(v_i)\\
0, otherwise
\end{cases},
\end{equation}
where $\mathcal{N}(v_i)$ denotes $k$ viewports nearest to the $i$-th viewport in the feature space. For simiplicity, we measure the distance between viewport $i$ and $p$ using the feature similarity:
\begin{equation}
s_{i,p} = \frac{x_i \cdot x_p}{max\{||x_i||_2\cdot ||x_p||_2, \epsilon\}},
\end{equation}
where $\epsilon$ is set as $10^{-12}$ to avoid division by zero.

\subsection{Viewport Quality Predictor}
Let us denote hierarchical representations of $N$ viewports as $\textbf{X} = \{x_1, x_2,..., x_N\} \in \mathbb{R}^{N\times md}$ and the hypergraph structure as $\textbf{E} = \{\textbf{E}_{loc}, \textbf{E}_{con} \} \in \mathbb{R}^{N\times 2N}$. The viewport quality predictor aims to derive the quality score of the distorted 360-degree image through hypergraph convolutional neural networks (HGCNs). Specifically, we first normalize the hypergraph structure $\textbf{E} $:
\begin{equation}
\hat{\textbf{E}} = \textbf{D}^{-1/2}_{v}\textbf{E}\textbf{D}^{-1}_{e}\textbf{E}^{T}\textbf{D}^{-1/2}_{v}
\end{equation}
where $\textbf{D}_{v} = diag\{\sum_j \textbf{E}_{1,j},..., \sum_j \textbf{E}_{N,j} \} \in \mathbb{R}^{N\times N}$ and $\textbf{D}_{e} = diag\{\sum_j \textbf{E}^T_{1,j},..., \sum_j \textbf{E}^T_{2N,j} \} \in \mathbb{R}^{2N\times 2N}$ are the node degree matrix and the edge degree matrix. Then, we conduct neighborhood aggregation using $n$ HGCN layers, which adopt the following layer-wise propagation rule:
\begin{equation}
\textbf{H}^{(t+1)} = \sigma(\text{BN}_{\gamma,\beta}(\hat{\textbf{E}}\textbf{H}^{(t)}\textbf{W}^{(t)}_1 + \textbf{H}^{(t)}\textbf{W}^{(t)}_2)), 
\end{equation}
where $\sigma$ is the Softplus activation function, $\textbf{H}^{(t+1)}$ is the output of the $t$-th HGCN layer, and $\textbf{H}^{(0)} = \textbf{X}$. $\textbf{W}^{(t)}_1$ and $\textbf{W}^{(t)}_2$ are learnable parameters. $\text{BN}_{\gamma,\beta}(\cdot)$ is the batch normalization with trainable parameters of $\gamma$ and $\beta$. Compared to the original version of HGCN~\cite{feng2019hypergraph}, we add the batch normalization and residual connection for accelerating and stablizing training. The final quality score is obtained by an average pooling layer:
\begin{equation}
Q = Mean(\textbf{H}^{n}),
\end{equation}
where $\textbf{H}^{n} \in \mathbb{R}^{N\times 1} $ denotes predicted quality scores of $N$ viewports. 
\begin{table}[ht]
	\begin{center}
		\captionsetup{justification=centering}
		\caption{\textsc{Details of the ResNet-18 architecture~\cite{resnet}.}}
		\label{table0}
		\begin{threeparttable}
			\begin{tabular}{@{}c|c|c@{}}
				\toprule
				Layer name                & Output size         & Layer                  \\ \midrule
				conv1                     & $112\times112\times64$      & $7\times7,\,64,$ stride 2 \\ \midrule
				\multirow{2}{*}{conv2\_x} & \multirow{2}{*}{$56\times56\times64$} & $3\times3$ max-pool, stride 2 \\
				& & $\left[ \begin{array}{l} 3 \times 3,\,64,\,1\\3 \times 3,\,64,\,1\end{array} \right]\times 2$  \\ \midrule
				conv3\_x  				  & $28\times28\times128$  
				& $\left[ \begin{array}{l} 3 \times 3,\,128,\,1\\3 \times 3,\,128,\,1\end{array} \right]\times 2$  \\ \midrule
				conv4\_x                  & $14\times14\times256$                  
				& $\left[ \begin{array}{l} 3 \times 3,\,256,\,1\\3 \times 3,\,256,\,1\end{array} \right]\times 2$  \\ \midrule
				conv5\_x                  & $7\times7\times512$                  
				& $\left[ \begin{array}{l} 3 \times 3,\,512,\,1\\3 \times 3,\,512,\,1\end{array} \right]\times 2$  \\ \midrule
				average pool& $1\times1\times512$              & $7\times7$ average pool \\\midrule
				fully connected& $1000$              & $512\times1000$ fully connections           \\ 	\midrule
				softmax & 1000 & \\			
				\bottomrule
			\end{tabular}
			\begin{tablenotes}
				\footnotesize
				\item[1] Convolutional layer: kernel size, channel, stride. 
				\item[2] Output size: height, width, channel.
			\end{tablenotes}
		\end{threeparttable}
	\end{center}
\end{table}
\subsection{Training Setup}
\noindent\textbf{Implementation Details:} For the viewport descriptor, following the liteature~\cite{xu2020blind}, we choose ResNet-18 architecture~\cite{resnet} as backbone network, and select features from conv2$\_5$,  conv$3\_9$, conv4$\_13$, and conv5$\_17$ layers. The details of the ResNet-18 network are presented  in Table~\ref{table0}. In the principle of ``\textit{reduce-pool-flatten-transform}",  each feature map is first converted  to a tensor with the size of $ 8 \times 8 \times 16 $ using convolutional neural networks with kernel size of 1 and the operation of \textit{MaxPooling2d}, which is later tranformed into a 256-dim vector by a fully connected (FC) layer. In constructing location-based hyperedges, we set the angular distance threshold as half of the viewport size, i.e., 45$\degree$. In constructing content-based hyperedges, we set $k$ as 5 and 0 for OIQA and CVIQD database through parameter tuning experiments. For the viewport quality predictor, similar to the configurations of GCN layers in the  literature~\cite{xu2020blind}, we use 5 HGCN layers and the viewport feature dimensions after each HGCN layer are [256, 128, 64, 32, 1]. 
\\ 
\noindent\textbf{Loss Function:} We use mean square error (MSE) as the training objective:
\begin{equation}
L_{mse} = \frac{1}{B} \sum_{b=1}^{B} (Q_b - G_b)^2, 
\end{equation}
where $B$ is the batch size, $Q_b$ and $G_b$ are the estimated and ground-truth MOS values of the $b$-th distorted 360-degree image.
\\ %
\noindent\textbf{Training Settings:} The backbone network is initialized by the pre-trained weights provided by the literature~\cite{xu2020blind}, while the viewport quality predictor is configured by the default initializer in Pytorch~\cite{paszke2019pytorch}. All trainable parameters are optimized by the Adam optimizer~\cite{kingma2014adam}. The learning rate of the backbone network is initialized as $1e^{-6}$ and fixed in the training phase, while that of the viewport quality predictor is initialized as $1e^{-3}$ and scaled by 0.25 every 40 epochs. The training epoch of the OIQA and CVIQD is set to 40 and 80.  During training, the mini-batch size is set to 16 and the input viewport is resized to $256 \times 256$. In addition, we use dropout~\cite{srivastava2014dropout} with a drop rate of 0.5 to avoid overfitting.

\section{Experiment}
\begin{table*}[htbp]
	\renewcommand\arraystretch{1.5}
	\begin{center}
		\captionsetup{justification=centering}
		\caption{\textsc{Performance comparison on OIQA database. The best FR and NR metrics are highlighted in bold.}}
		\label{table1}
		\scalebox{0.8}{
			\begin{tabular}{@{}cc|ccc|ccc|ccc|ccc|ccc@{}}
				\toprule
				&              & \multicolumn{3}{c|}{JPEG}                           & \multicolumn{3}{c|}{JP2K}                           & \multicolumn{3}{c|}{WN}                             & \multicolumn{3}{c|}{BLUR}                           & \multicolumn{3}{c}{ALL}                             \\ \midrule
				&              & PLCC            & SROCC           & RMSE            & PLCC            & SROCC           & RMSE            & PLCC            & SROCC           & RMSE            & PLCC            & SROCC           & RMSE            & PLCC            & SROCC           & RMSE            \\
				\multirow{8}{*}{FR} & PSNR         & 0.6941          & 0.7060          & 1.6141          & 0.8632          & 0.7821          & 1.1316          & 0.9547          & 0.9500          & 0.5370          & 0.9282          & 0.7417          & 0.8299          & 0.5812          & 0.5226          & 1.7005          \\
				& S-PSNR \cite{SPSNR}& 0.6911          & 0.6148          & 1.6205          & 0.9205          & 0.7250          & 0.8757          & 0.9503          & 0.9357          & 0.5620          & 0.8282          & 0.7525          & 1.0910          & 0.5997          & 0.5399          & 1.6721          \\
				& WS-PSNR \cite{WSPSNR}& 0.7133          & 0.6792          & 1.5713          & 0.9344          & 0.7500          & 0.9128          & 0.9626          & 0.9500          & 0.4890          & 0.8190          & 0.7668          & 1.1172          & 0.5819          & 0.5263          & 1.6994          \\
				& CPP-PSNR \cite{CPPPSNR}& 0.6153          & 0.5362          & 1.7693          & 0.8971          & 0.7250          & 0.9904          & 0.9276          & 0.9143          & 0.6739          & 0.7969          & 0.7185          & 1.1728          & 0.5683          & 0.5149          & 1.7193          \\
				& SSIM \cite{ssim}& 0.9077          & \textbf{0.9008} & 0.9406          & 0.9783          & 0.9679          & 0.4643          & 0.8828          & 0.8607          & 0.8474          & 0.9926          & 0.9777          & 0.2358          & 0.8718          & 0.8588          & 1.0238          \\
				& MS-SSIM \cite{msssim}& \textbf{0.9102} & 0.8937          & \textbf{0.9288} & 0.9492          & 0.9250          & 0.7052          & 0.9691          & 0.9571          & 0.4452          & 0.9251          & 0.8990          & 0.7374          & 0.7710          & 0.7379          & 1.3308          \\
				& FSIM \cite{fsim}& 0.8938          & 0.8490          & 1.0057          & 0.9699          & 0.9643          & 0.5454          & 0.9170          & 0.8893          & 0.7197          & \textbf{0.9914} & \textbf{0.9902} & \textbf{0.2544} & 0.9014          & 0.8938          & 0.9047          \\
				& DeepQA \cite{DeepQA}& 0.8301          & 0.8150          & 1.2506          & \textbf{0.9905} & \textbf{0.9893} & \textbf{0.3082} & \textbf{0.9709} & \textbf{0.9857} & \textbf{0.4317} & 0.9623          & 0.9473          & 0.5283          & \textbf{0.9044} & \textbf{0.8973} & \textbf{0.8914} \\ \midrule
				\multirow{6}{*}{NR} & BRISQUE \cite{BRISQUE}& 0.9160          & 0.9392          & 0.8992          & 0.7397          & 0.6750          & 1.5082          & \textbf{0.9818}          & 0.9750          & \textbf{0.3427}          & 0.8663          & 0.8508          & 0.9697          & 0.8424          & 0.8331          & 1.1261          \\
				& BMPRI \cite{BMPRI}& 0.9361          & 0.8954          & 0.7886          & 0.8322          & 0.8214          & 1.2428          & 0.9673          & 0.9821          & 0.4572          & 0.5199          & 0.3807          & 1.6584          & 0.6503          & 0.6238          & 1.5874          \\
				& DB-CNN \cite{DBCNN}& 0.8413          & 0.7346          & 1.2118          & 0.9755          & {0.9607} & 0.4935          & 0.9772 & \textbf{0.9786} & 0.3832 & 0.9536          & 0.8865          & 0.5875          & 0.8852          & 0.8653          & 0.9717          \\
				& MC360IQA \cite{mc360iqa2}& 0.9459          & 0.9008          & 0.7272          & 0.9165          & 0.9036          & 0.8966          & 0.9718          & 0.9464          & 0.4251          & 0.9526          & 0.9580          & 0.5907          & 0.9267          & 0.9139          & 0.7854         \\
				& VGCN (local) \cite{xu2020blind} & 0.9508          & 0.8972          & 0.6949          & {0.9793} & 0.9439          & {0.4541} & 0.9682          & 0.9714          & 0.4515          & \textbf{0.9838}          & \textbf{0.9759} & \textbf{0.3479 }         & 0.9529          & 0.9444          & 0.6340          \\
				& AHGCN         & \textbf{0.9649} & \textbf{0.9276} & \textbf{0.5886} & \textbf{0.9820 }         & \textbf{0.9643   }       & \textbf{0.4236 }         & 0.9706         & \textbf{0.9786 }      & 0.4341         & {0.9756} & \textbf{0.9759 }       & {0.4264} & \textbf{0.9649} & \textbf{0.9590} & \textbf{0.5487} \\ 				
				\bottomrule
		\end{tabular}}
	\end{center}
\end{table*}

\begin{figure*}[h]
	\centering
	\subfigure[PSNR]{
		\includegraphics[height=2.4cm]{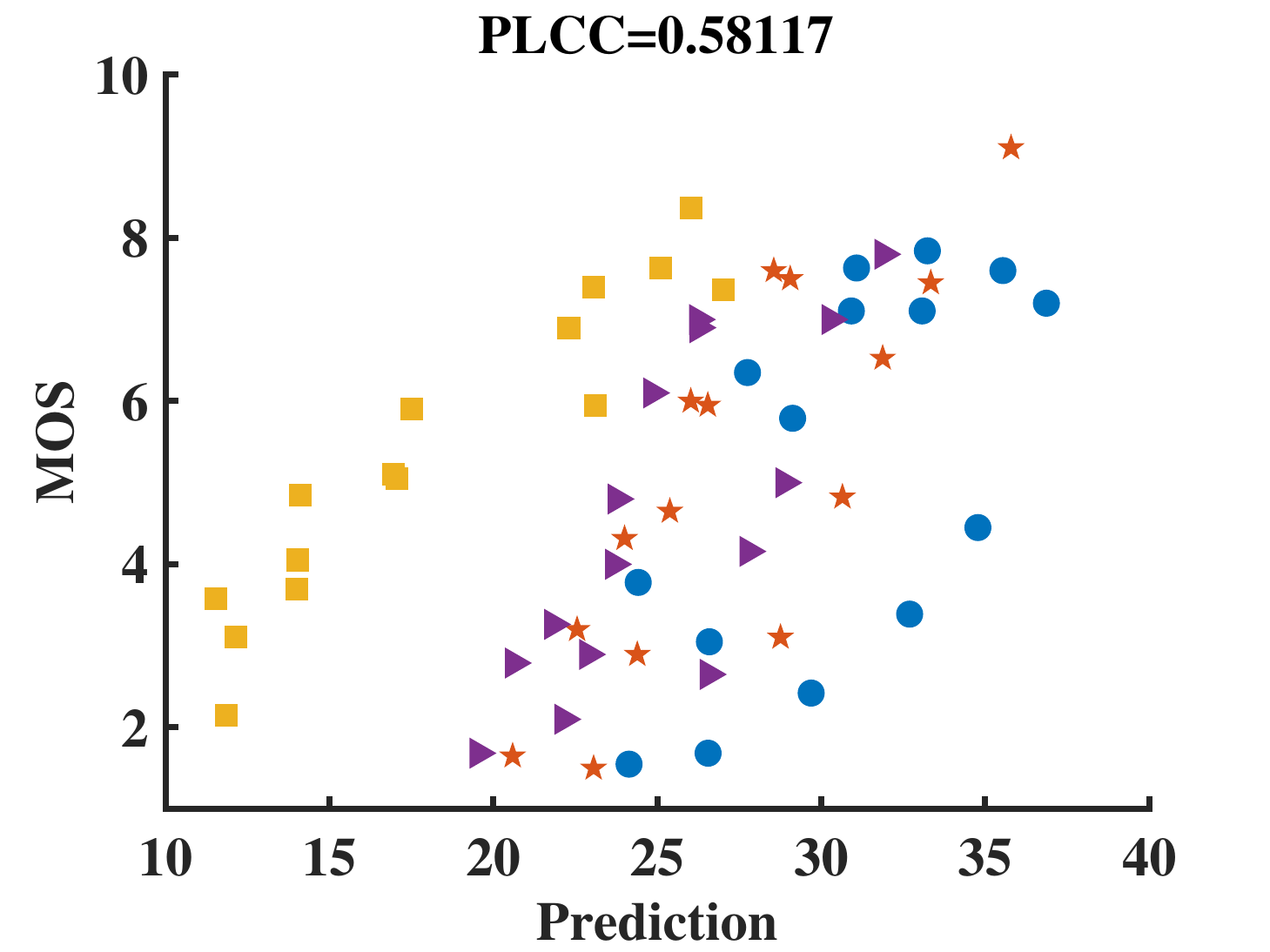}
	}
	\subfigure[S-PSNR]{
		\includegraphics[height=2.4cm]{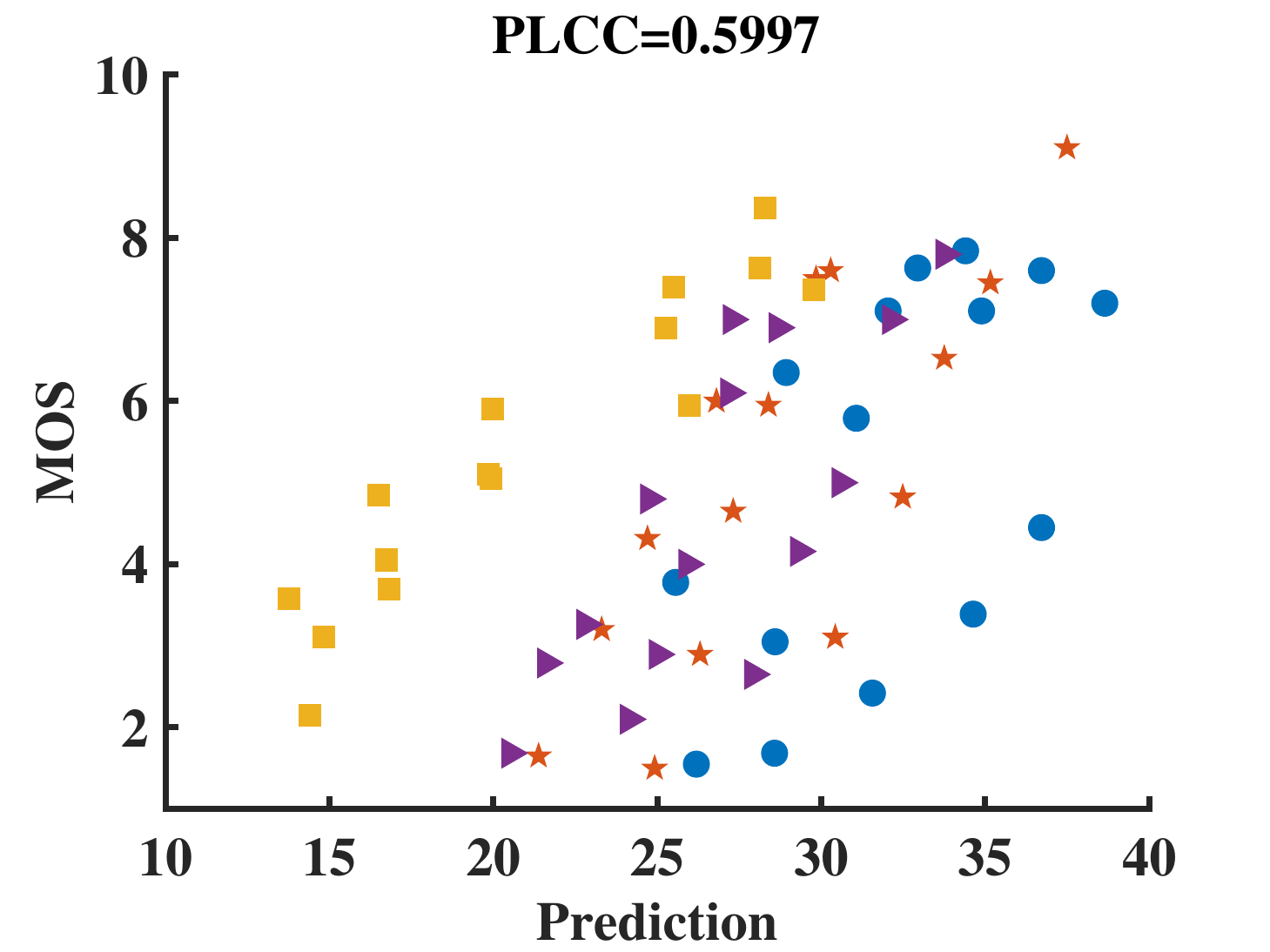}
	}
	\subfigure[WS-PSNR]{
		\includegraphics[height=2.4cm]{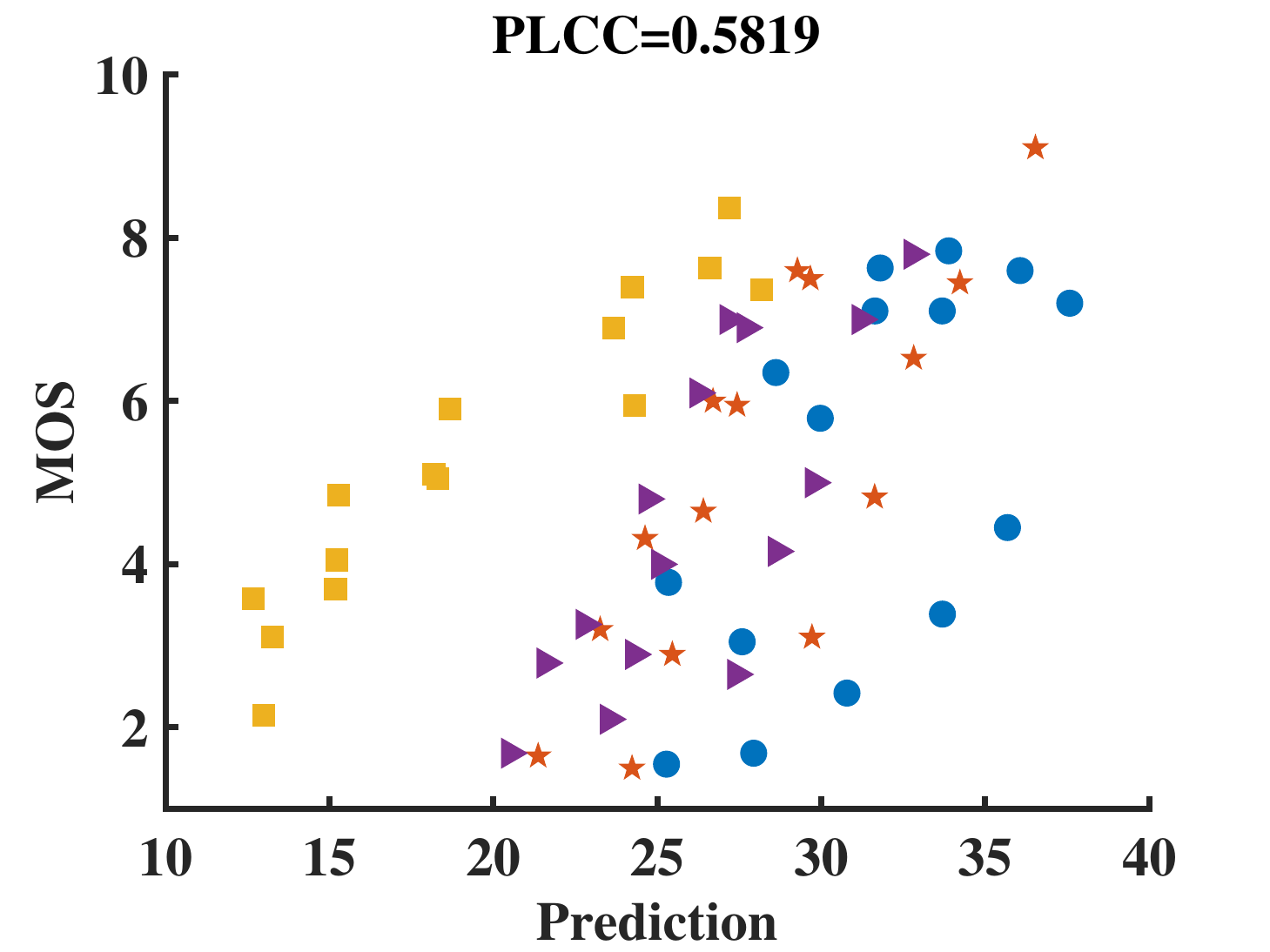}
	}
	\subfigure[CPP-PSNR]{
		\includegraphics[height=2.4cm]{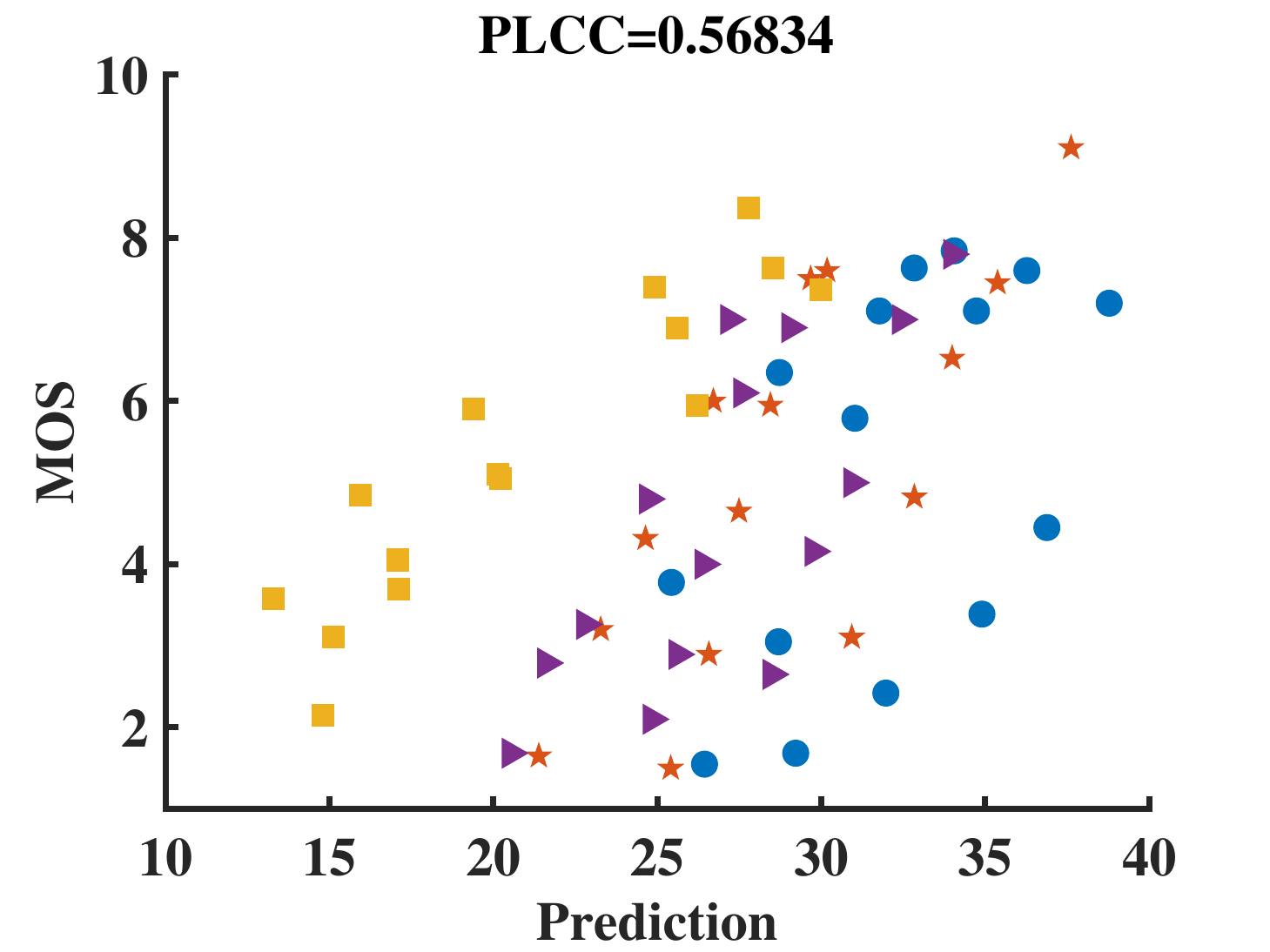}
	}
	\subfigure[SSIM]{
		\includegraphics[height=2.4cm]{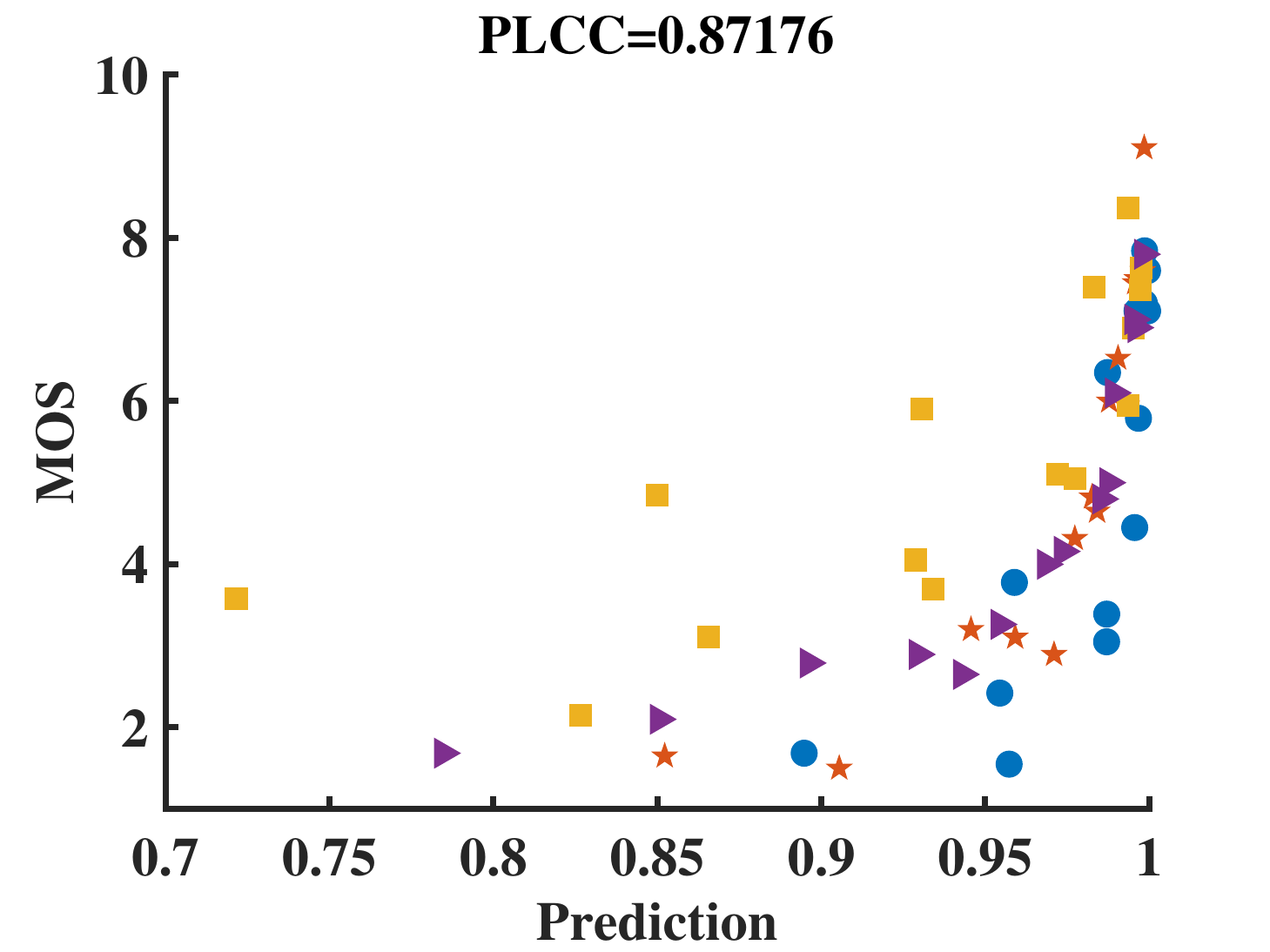}
	}
	\subfigure[MS-SSIM]{
		\includegraphics[height=2.4cm]{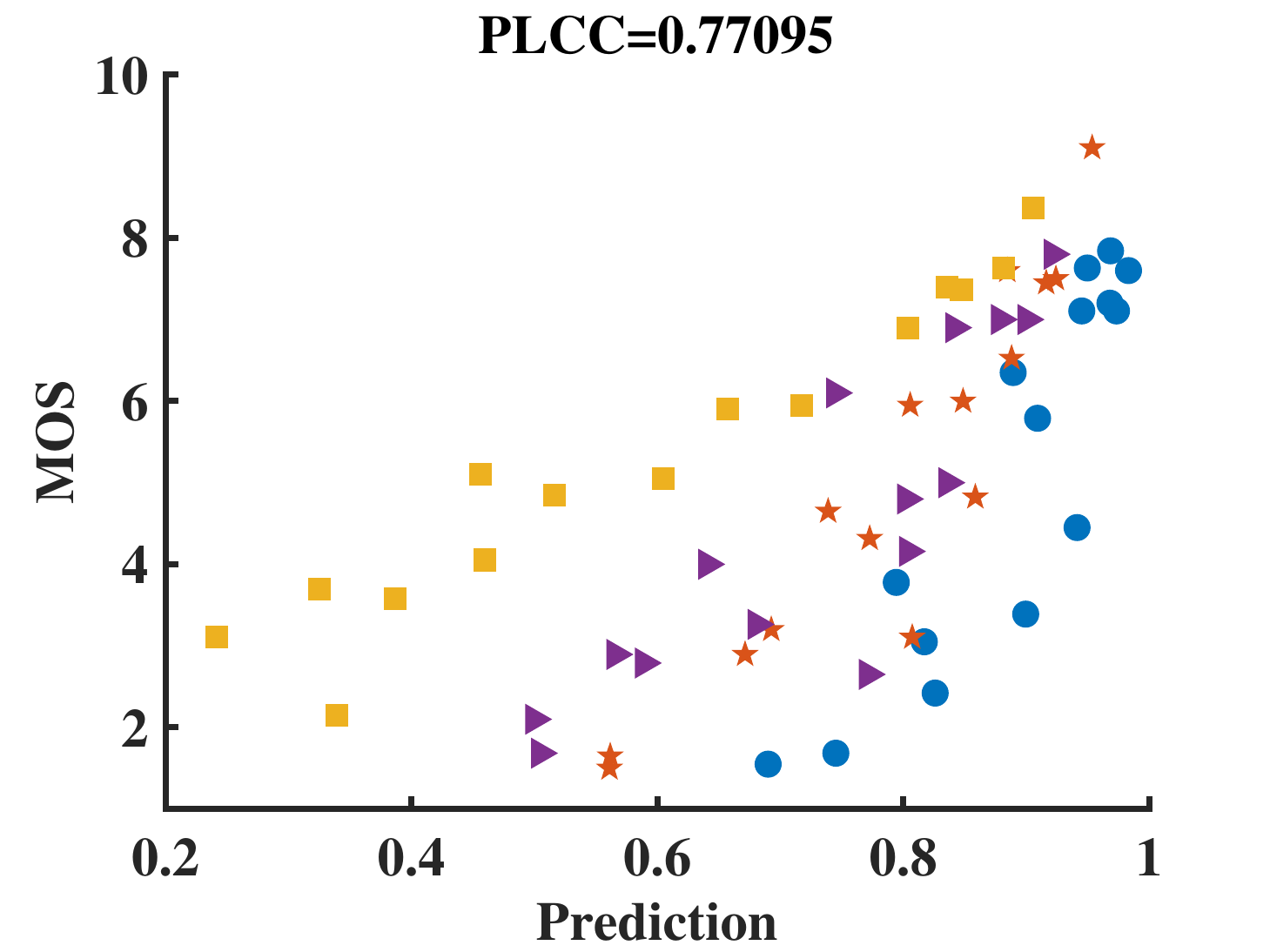}
	}
	\subfigure[FSIM]{
		\includegraphics[height=2.4cm]{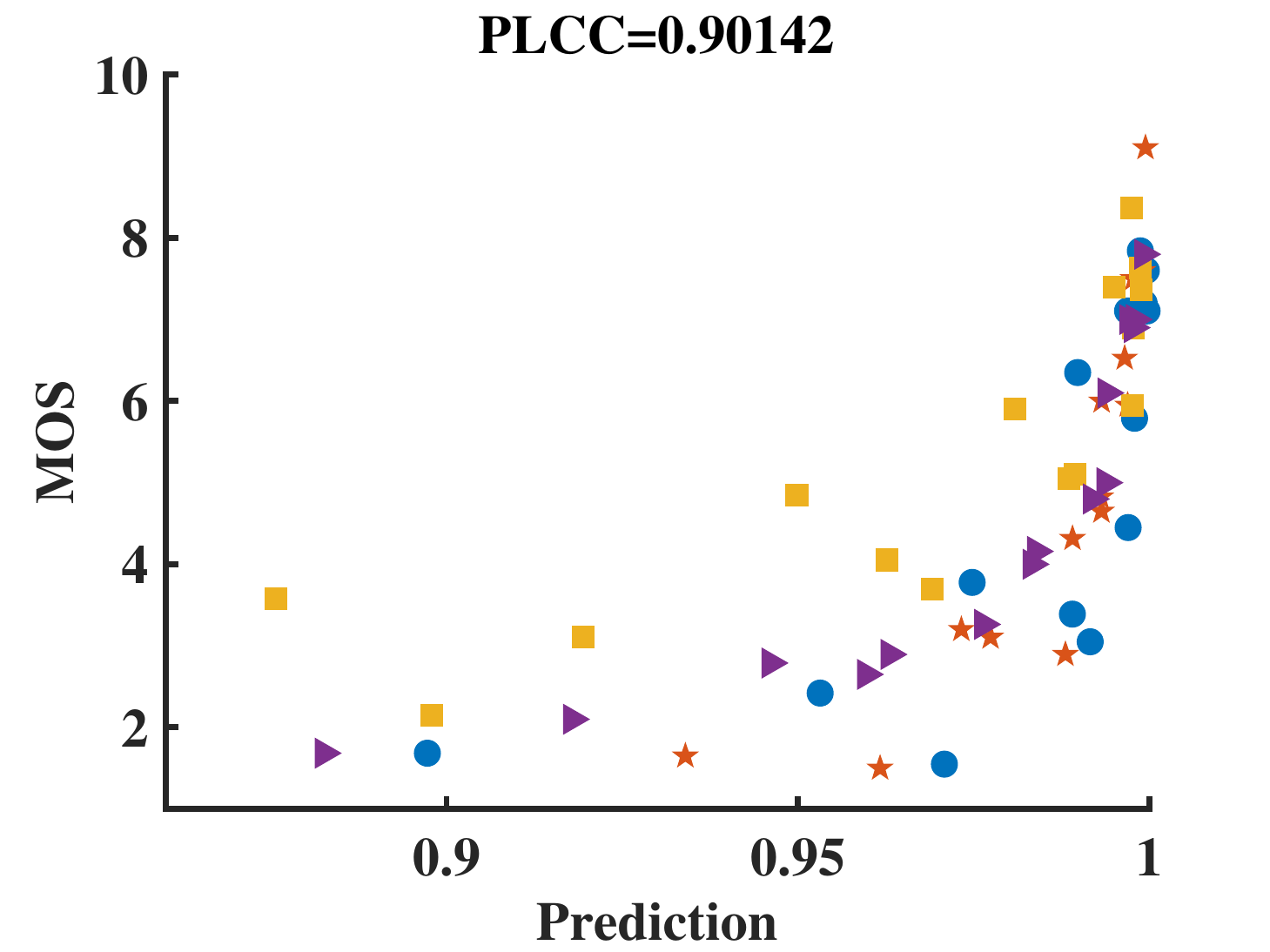}
	}
	\subfigure[DeepQA]{
		\includegraphics[height=2.4cm]{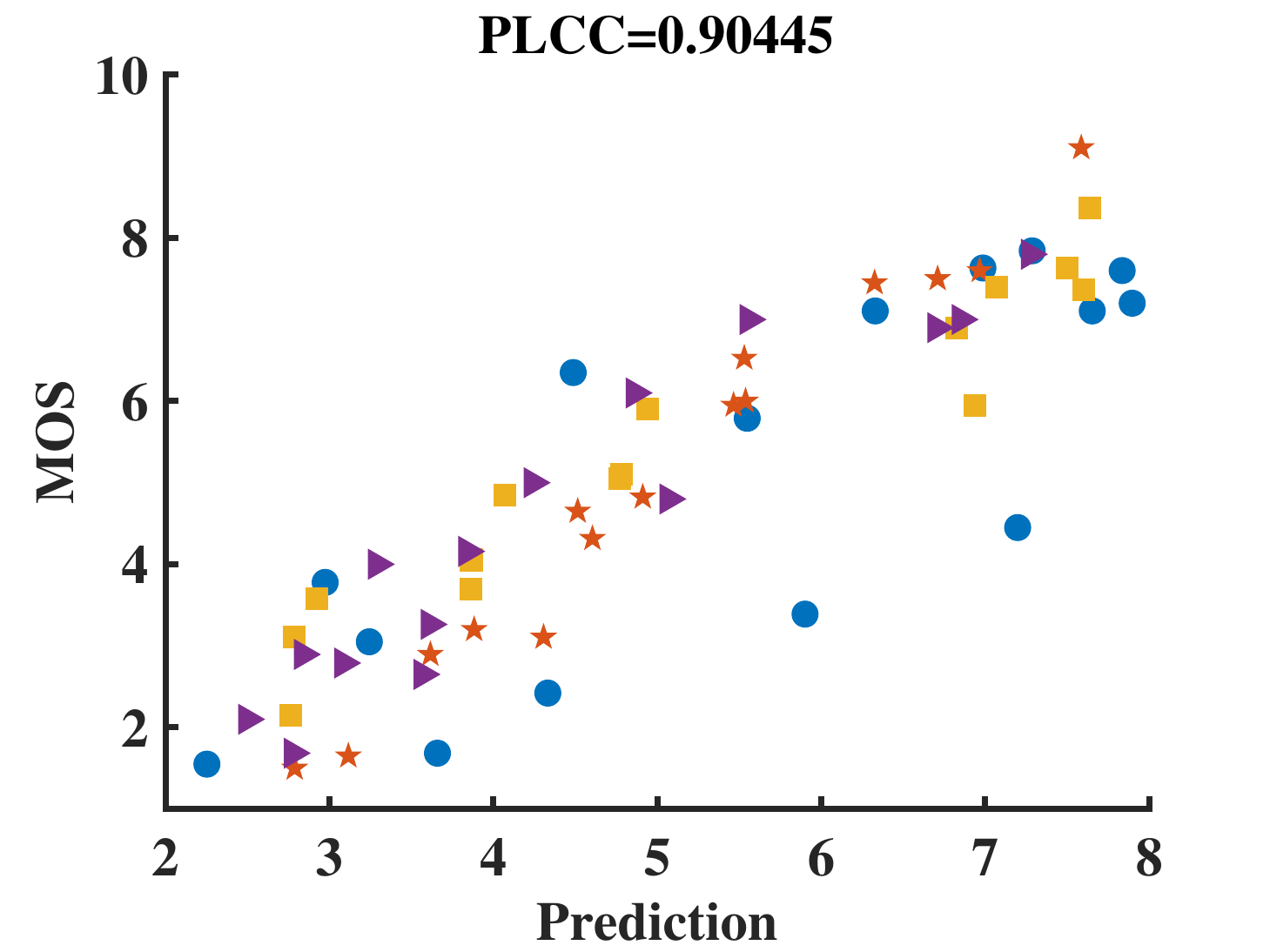}
	}
	\subfigure[BRISQUE]{
		\includegraphics[height=2.4cm]{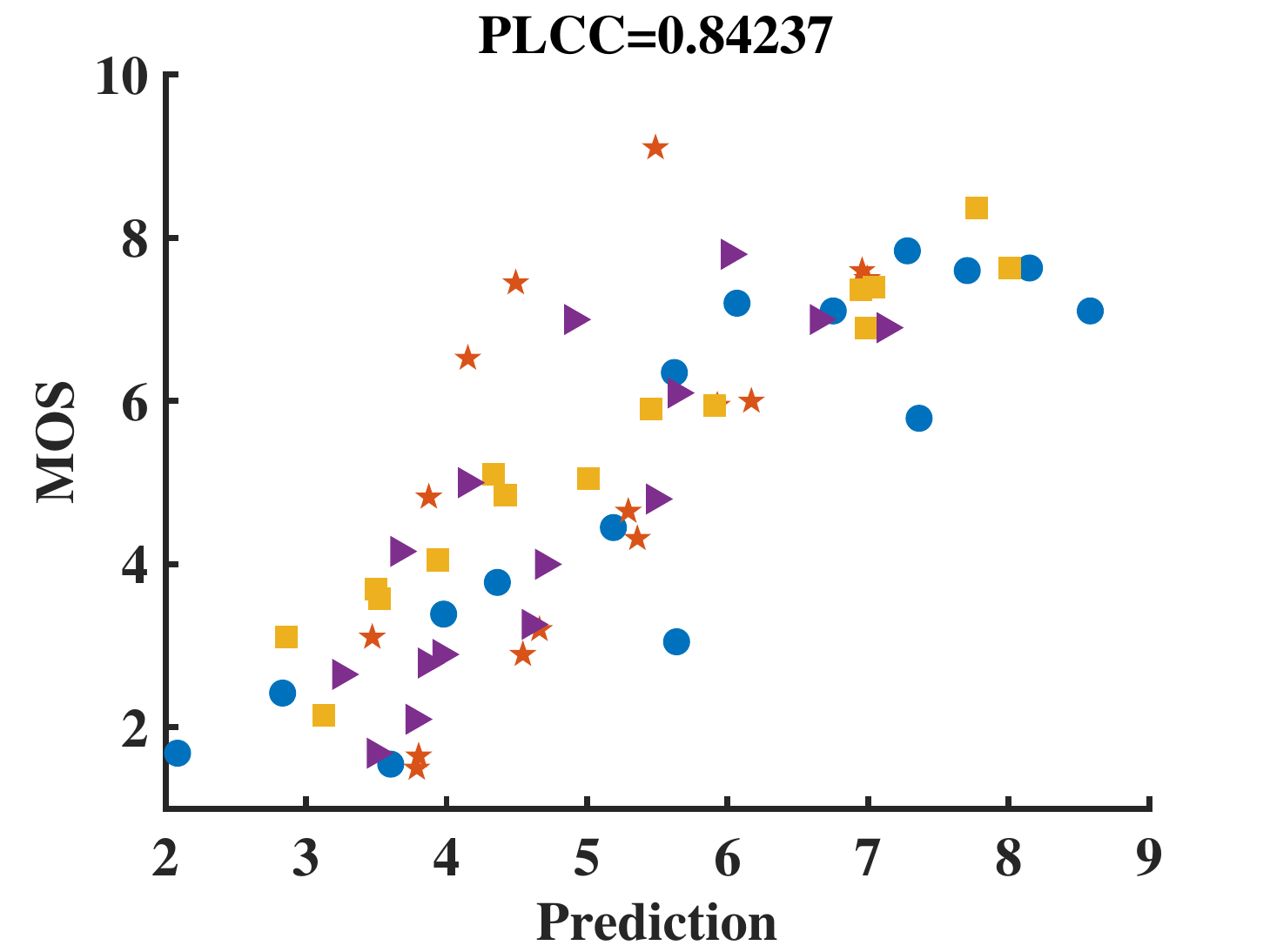}
	}
	\subfigure[BMPRI]{
		\includegraphics[height=2.4cm]{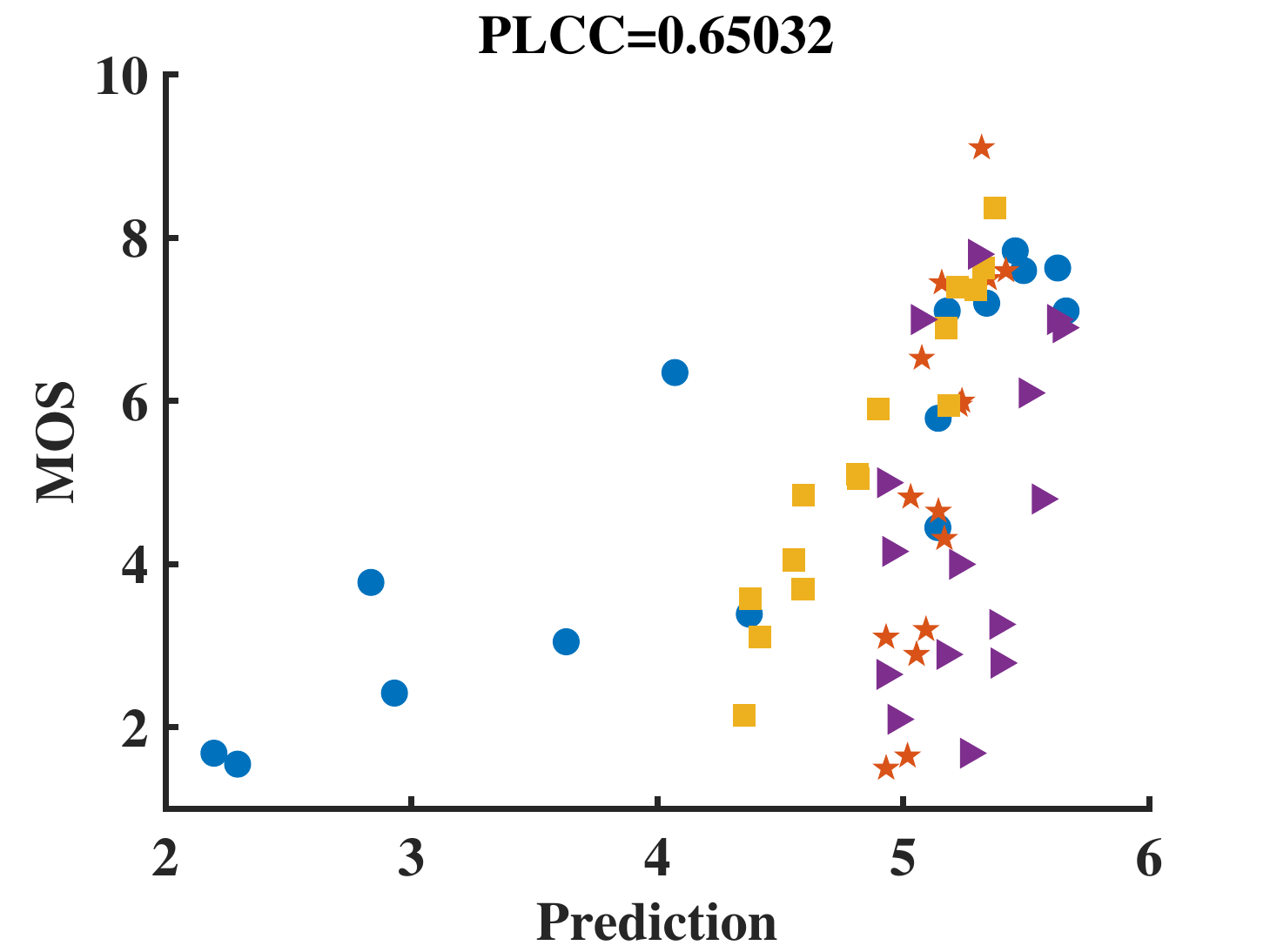}
	}
	\subfigure[DB-CNN]{
		\includegraphics[height=2.4cm]{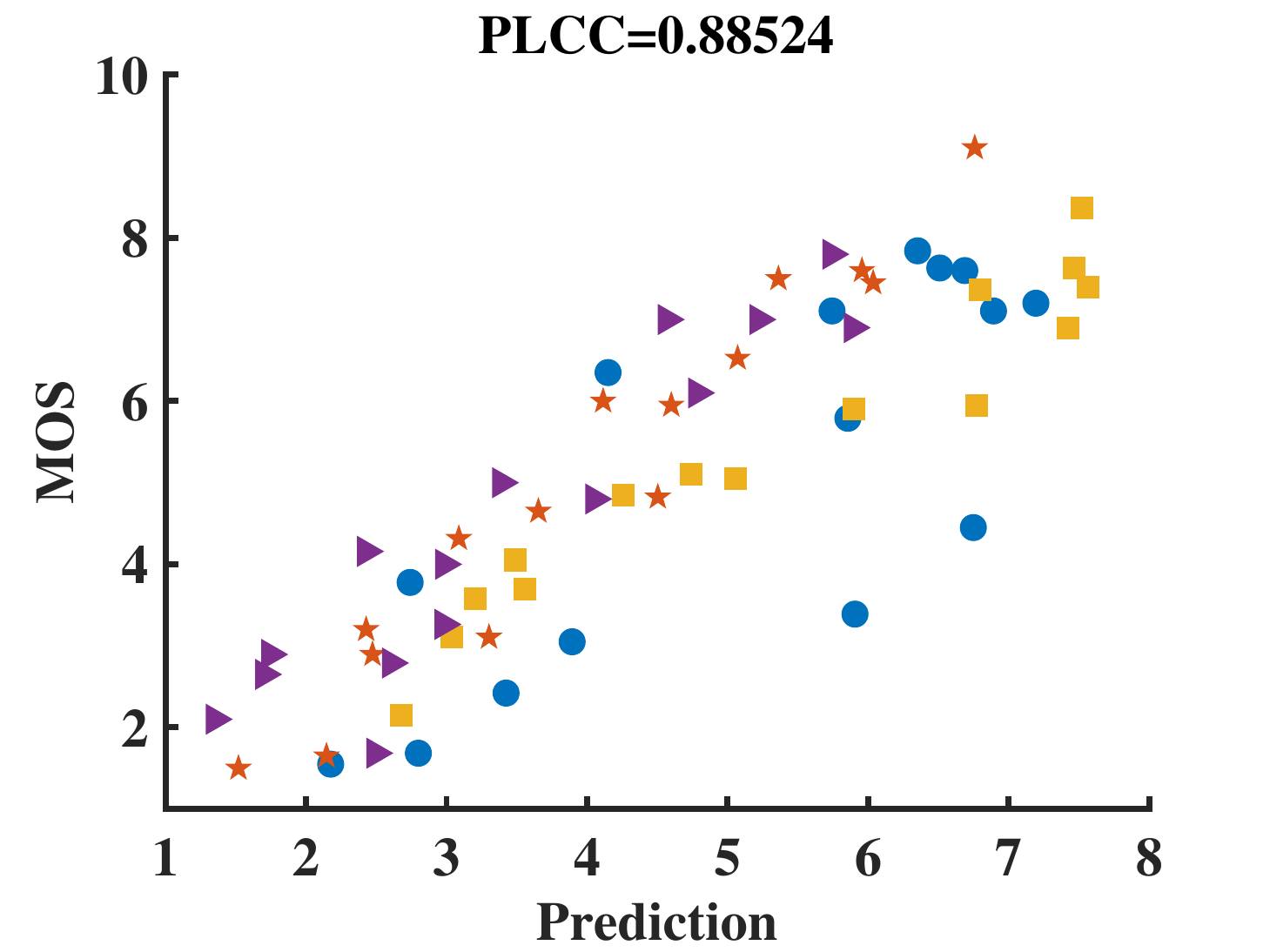}
	}
	\subfigure[MC360IQA]{
		\includegraphics[height=2.4cm]{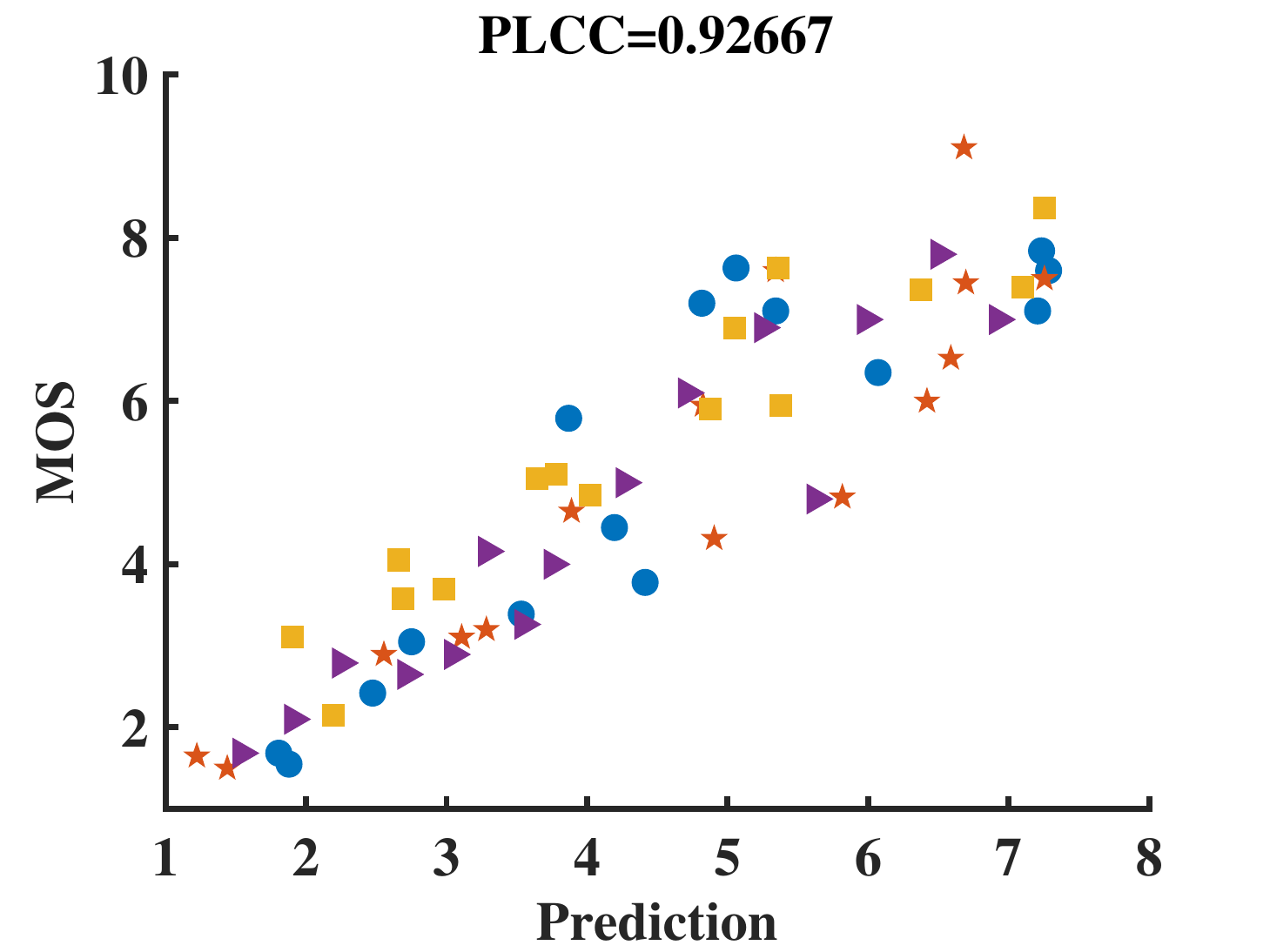}
	}
	\subfigure[VGCN (local)]{
		\includegraphics[height=2.4cm]{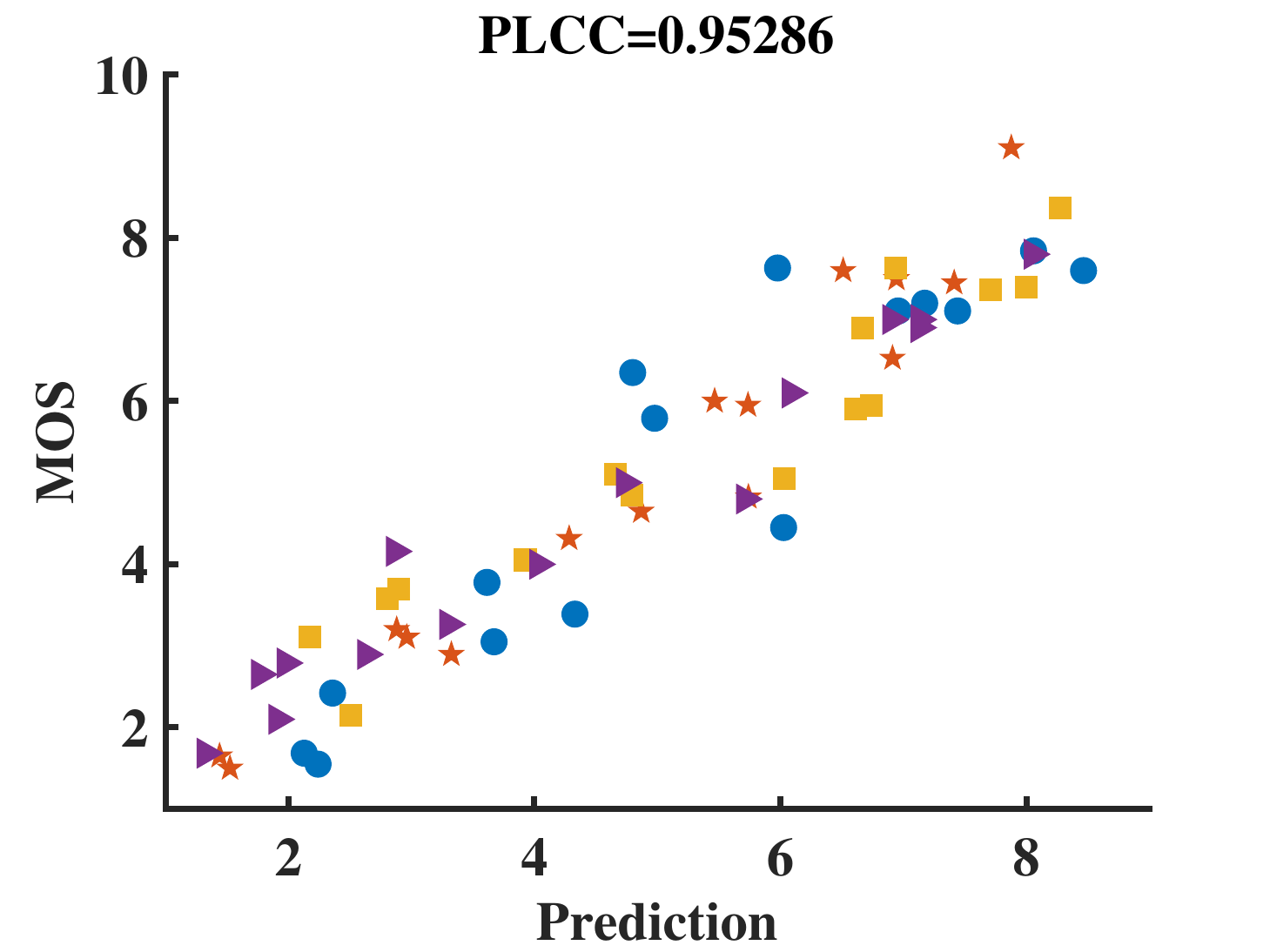}
	}
	\subfigure[AHGCN]{
		\includegraphics[height=2.4cm]{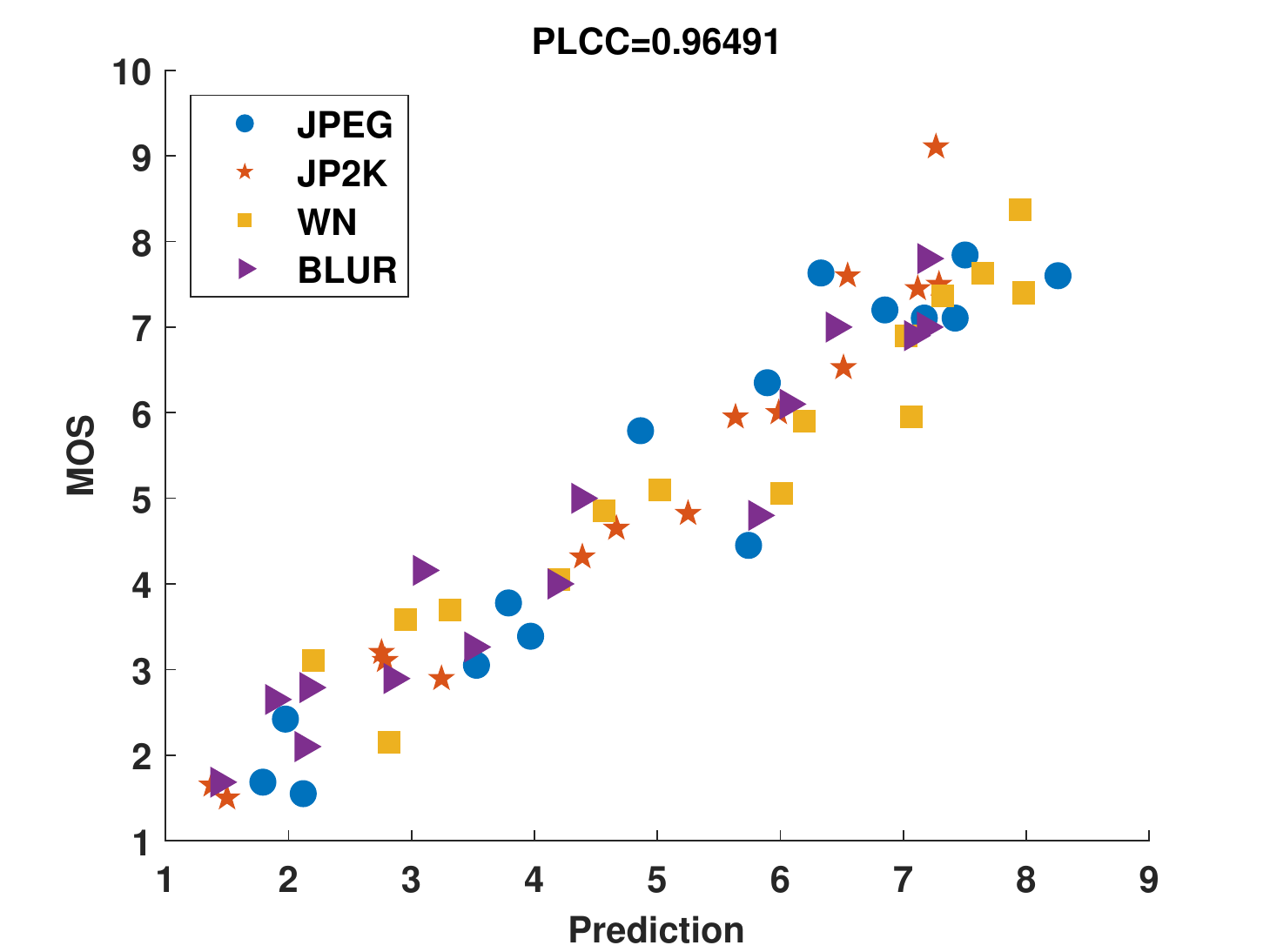}
	}
	\caption{Scatter plots of MOS values versus predictions of IQA metrics on the testing set of OIQA database.}
	\label{fig:fig4}
\end{figure*}
\begin{table*}[htbp]
	\renewcommand\arraystretch{1.5}
	\begin{center}
		\caption{\textsc{Performance comparison on CVIQD database. The best FR and NR metrics are highlighted in bold.}}
		\label{table2}
		\scalebox{0.8}{
			\begin{tabular}{@{}cc|ccc|ccc|ccc|ccc@{}}
				\toprule
				&              & \multicolumn{3}{c|}{JPEG}                            & \multicolumn{3}{c|}{AVC}                             & \multicolumn{3}{c|}{HEVC}                            & \multicolumn{3}{c}{ALL}                             \\ \midrule
				&              & PLCC            & SROCC           & RMSE            & PLCC            & SROCC           & RMSE            & PLCC            & SROCC           & RMSE            & PLCC            & SROCC           & RMSE            \\
				\multirow{8}{*}{FR} & PSNR         & 0.8682          & 0.6982          & 8.0429          & 0.6141          & 0.5802          & 10.5520         & 0.5982          & 0.5762          & 9.4697          & 0.7008          & 0.6239          & 9.9599          \\
				& S-PSNR \cite{SPSNR}      & 0.8661          & 0.7172          & 8.1008          & 0.6307          & 0.6039          & 10.3760         & 0.6514          & 0.6150          & 8.9585          & 0.7083          & 0.6449          & 9.8564          \\
				& WS-PSNR \cite{WSPSNR}     & 0.8572          & 0.6848          & 8.3465          & 0.5702          & 0.5521          & 10.9841         & 0.5884          & 0.5642          & 9.5473          & 0.6729          & 0.6107          & 10.3283         \\
				& CPP-PSNR \cite{CPPCNR}    & 0.8585          & 0.7059          & 8.3109          & 0.6137          & 0.5872          & 10.5615         & 0.6160          & 0.5689          & 9.3009          & 0.6871          & 0.6265          & 10.1448         \\
				& SSIM \cite{ssim}        & 0.9822          & 0.9582          & 3.0468          & 0.9303          & 0.9174          & 4.9029          & 0.9436          & 0.9452          & 3.9097          & 0.9002          & 0.8842          & 6.0793          \\
				& MS-SSIM \cite{msssim}     & 0.9636          & 0.9047          & 4.3355          & 0.7960          & 0.7650          & 8.0924          & 0.8072          & 0.8011          & 6.9693          & 0.8521          & 0.8222          & 7.3072          \\
				& FSIM \cite{fsim}        & \textbf{0.9839} & \textbf{0.9639} & \textbf{2.8928} & \textbf{0.9534} & \textbf{0.9439} & \textbf{4.0327} & \textbf{0.9617} & \textbf{0.9532} & \textbf{3.2385} & 0.9340          & 0.9152          & 4.9864          \\
				& DeepQA \cite{DeepQA}      & 0.9526          & 0.9001          & 4.9290          & 0.9477          & 0.9375          & 4.2683          & 0.9221          & 0.9288          & 4.5694          & \textbf{0.9375} & \textbf{0.9292} & \textbf{4.8574} \\ \midrule
				\multirow{6}{*}{NR} & BRISQUE \cite{BRISQUE}     & 0.9464          & 0.9031          & 5.2442          & 0.7745          & 0.7714          & 8.4573          & 0.7548          & 0.7644          & 7.7455          & 0.8376          & 0.8180          & 7.6271          \\
				& BMPRI \cite{BMPRI}       & 0.9874          & 0.9562          & 2.5597          & 0.7161          & 0.6731          & 9.3318          & 0.6154          & 0.6715          & 9.3071          & 0.7919          & 0.7470          & 8.5258          \\
				& DB-CNN \cite{DBCNN}       & 0.9779          & 0.9576          & 3.3862          & 0.9564          & 0.9545          & 3.9063          & 0.8646          & 0.8693          & 5.9335          & 0.9356          & 0.9308          & 4.9311          \\
				& MC360IQA \cite{mc360iqa2}    & 0.9698          & 0.9693          & 3.9517          & 0.9487          & 0.9569          & 4.2281          & 0.8976          & 0.9104          & 5.2557          & 0.9429          & 0.9428          & 4.6506          \\
				& VGCN (local) \cite{xu2020blind} & 0.9857          & 0.9666          & 2.7310          & 0.9684          & 0.9622          & 3.3328          & 0.9367          & \textbf{0.9422 }         & 4.1329          & 0.9597          & 0.9539          & 3.9220          \\
				& AHGCN       & \textbf{0.9869} & \textbf{0.9686} & \textbf{2.6162} & \textbf{0.9793} & \textbf{0.9753} & \textbf{2.7084} & \textbf{0.9419} & {0.9412} & \textbf{3.9657} & \textbf{0.9643} & \textbf{0.9623} & \textbf{3.6990} \\ 				
				
				\bottomrule
		\end{tabular}}
	\end{center}
\end{table*}

\begin{figure*}[h]
	\centering
		\subfigure[PSNR]{
			\includegraphics[height=2.4cm]{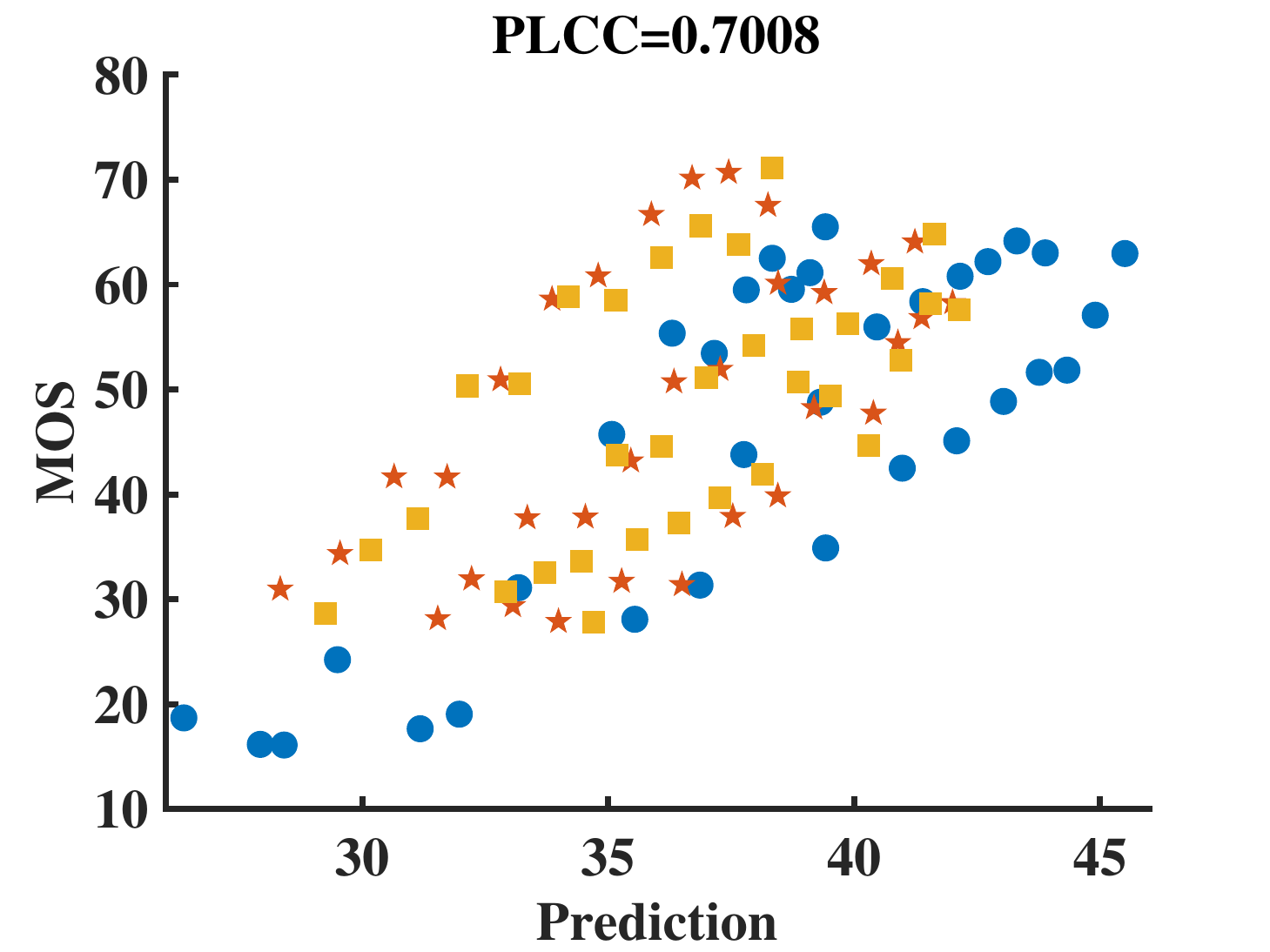}
		}
		\subfigure[S-PSNR]{
			\includegraphics[height=2.4cm]{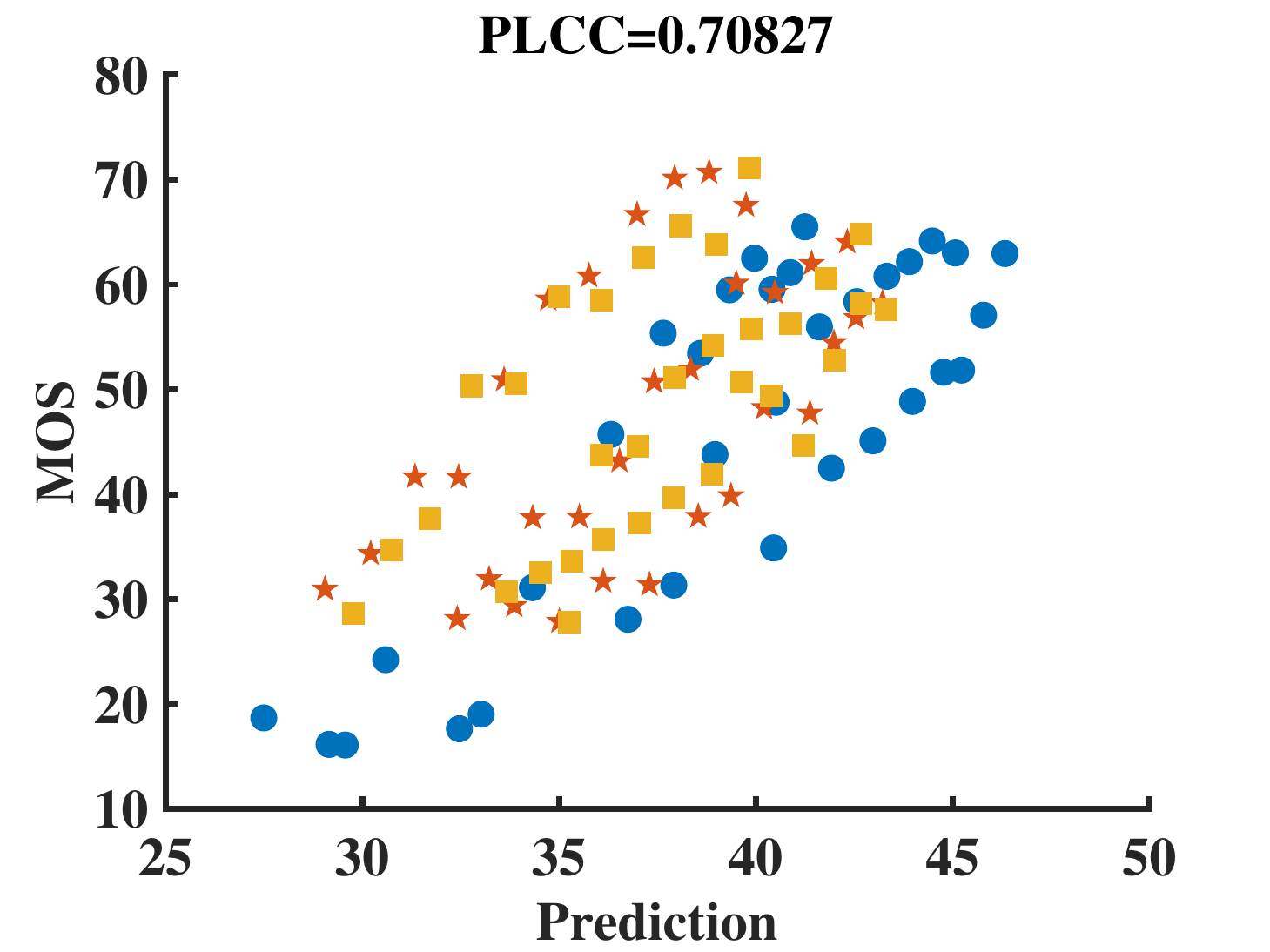}
		}
	\subfigure[WS-PSNR]{
		\includegraphics[height=2.4cm]{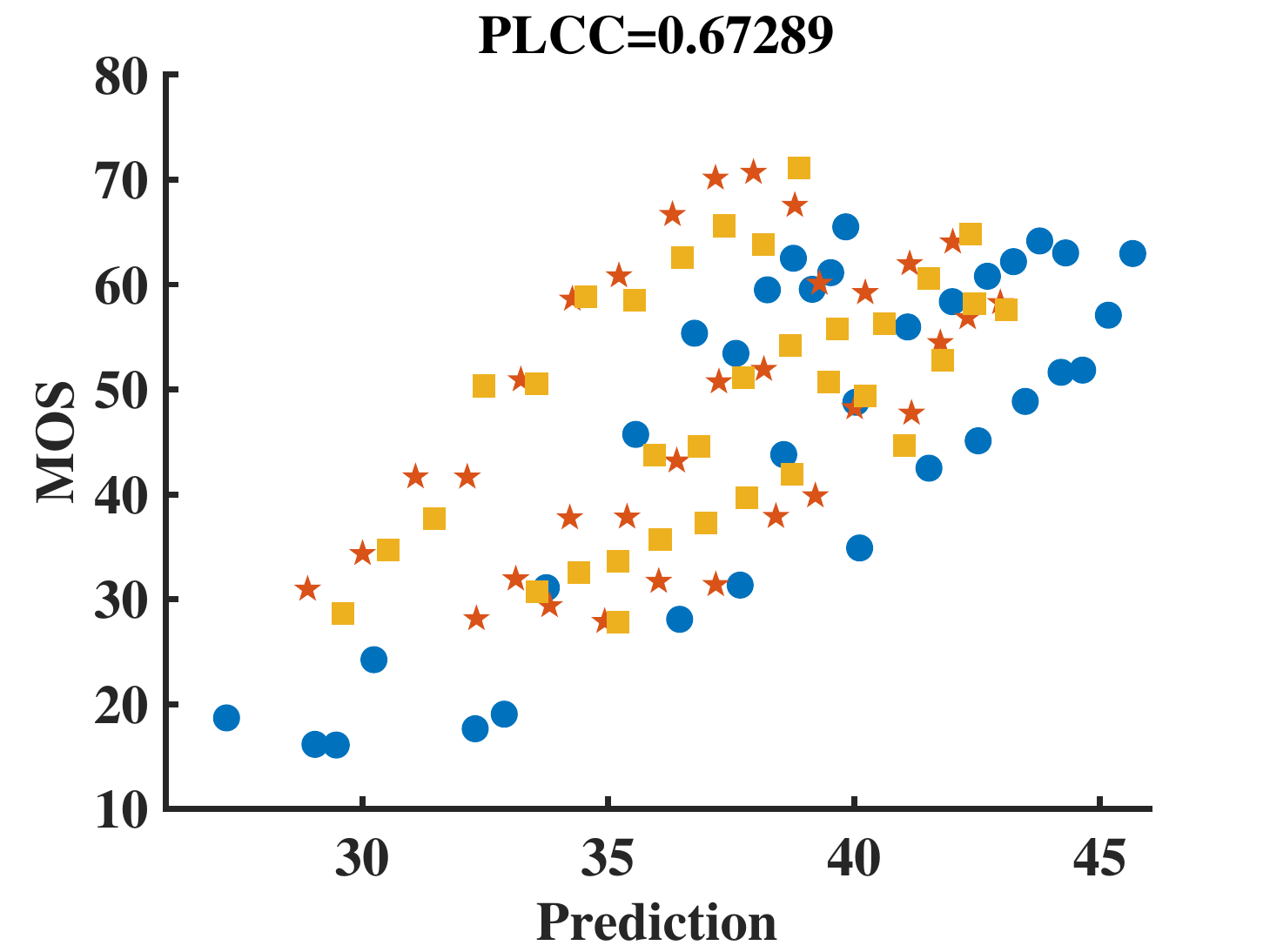}
	}
		\subfigure[CPP-PSNR]{
			\includegraphics[height=2.4cm]{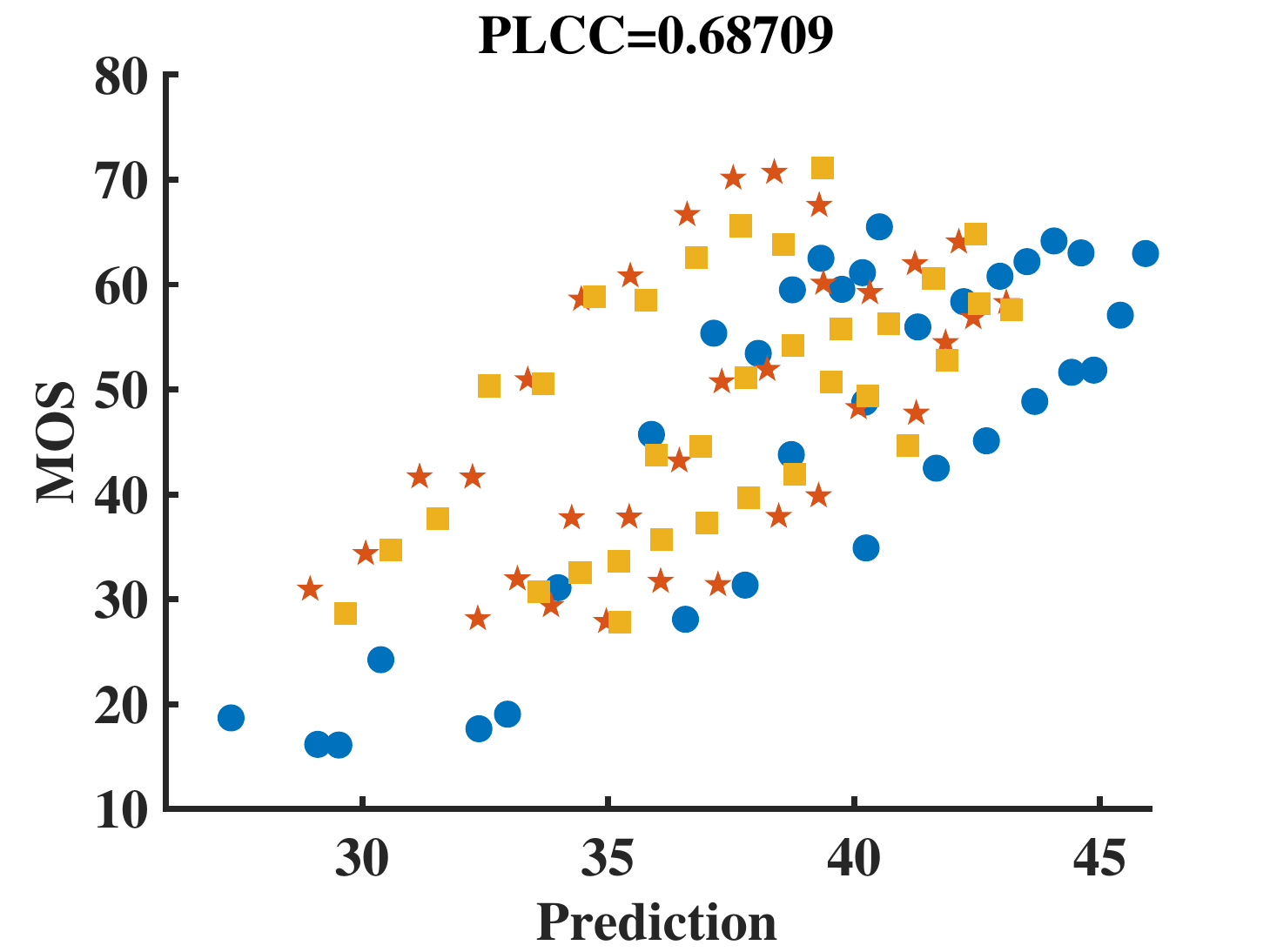}
		}
	\subfigure[SSIM]{
		\includegraphics[height=2.4cm]{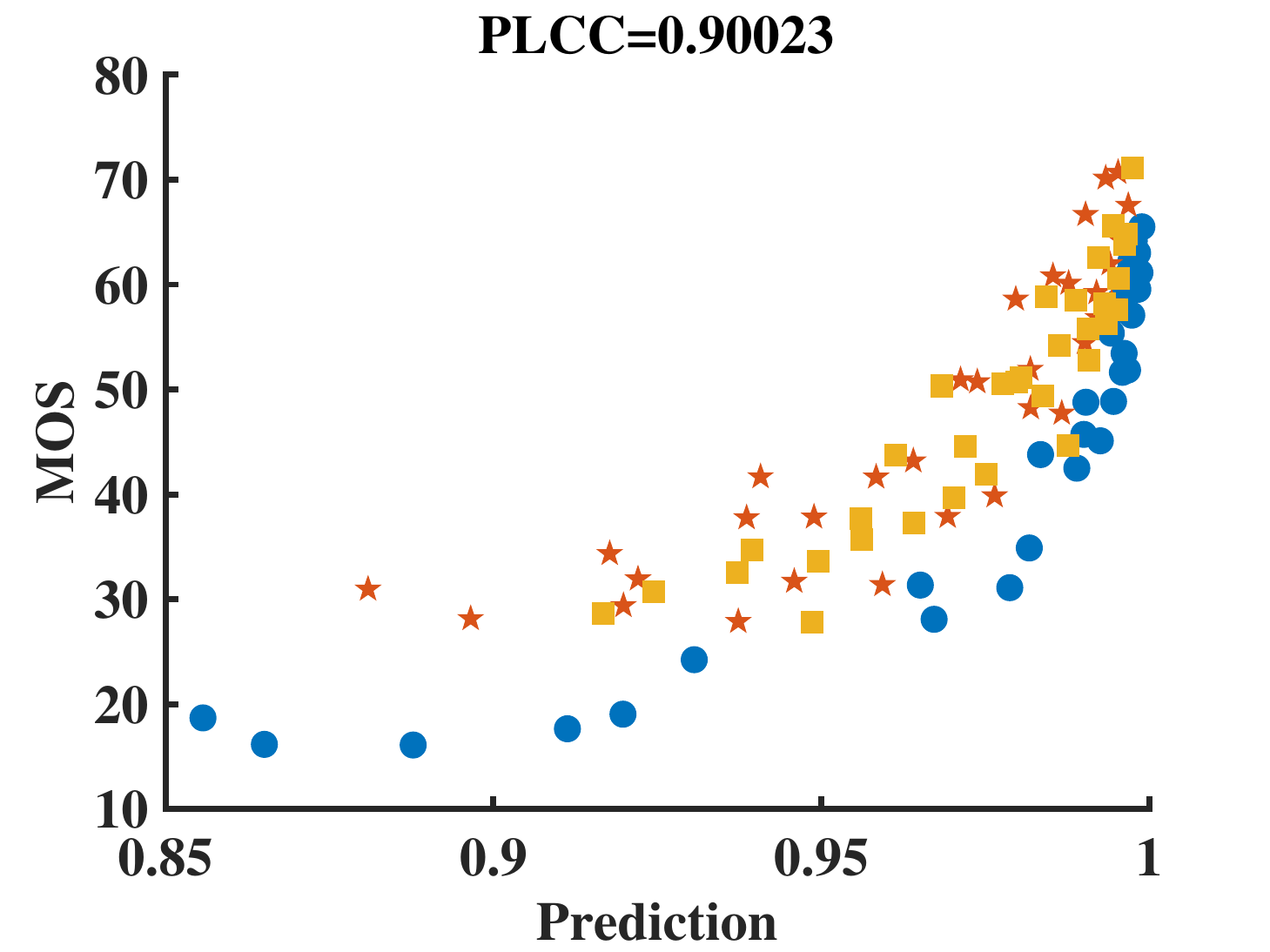}
	}
		\subfigure[MS-SSIM]{
			\includegraphics[height=2.4cm]{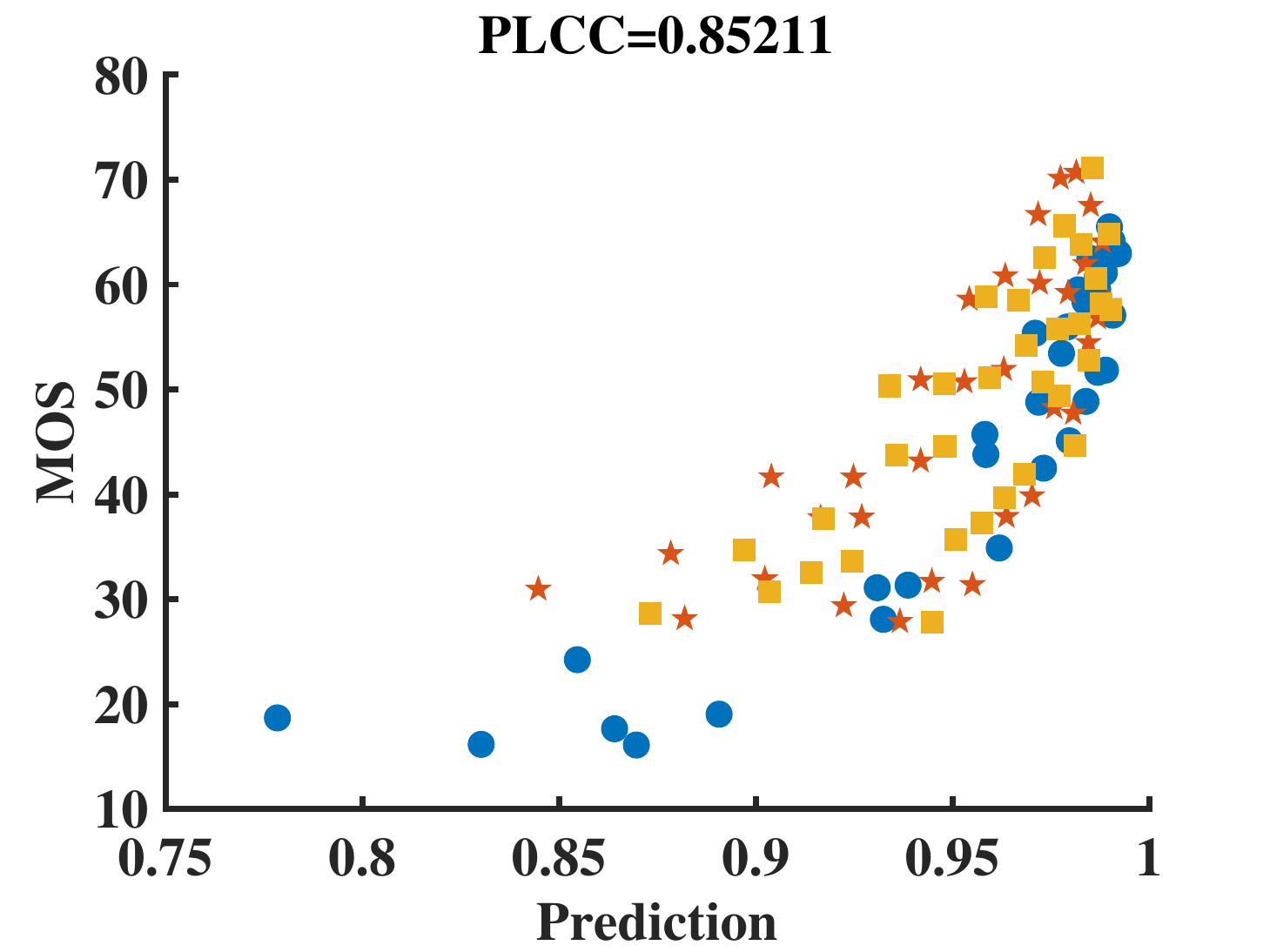}
		}
	\subfigure[FSIM]{
		\includegraphics[height=2.4cm]{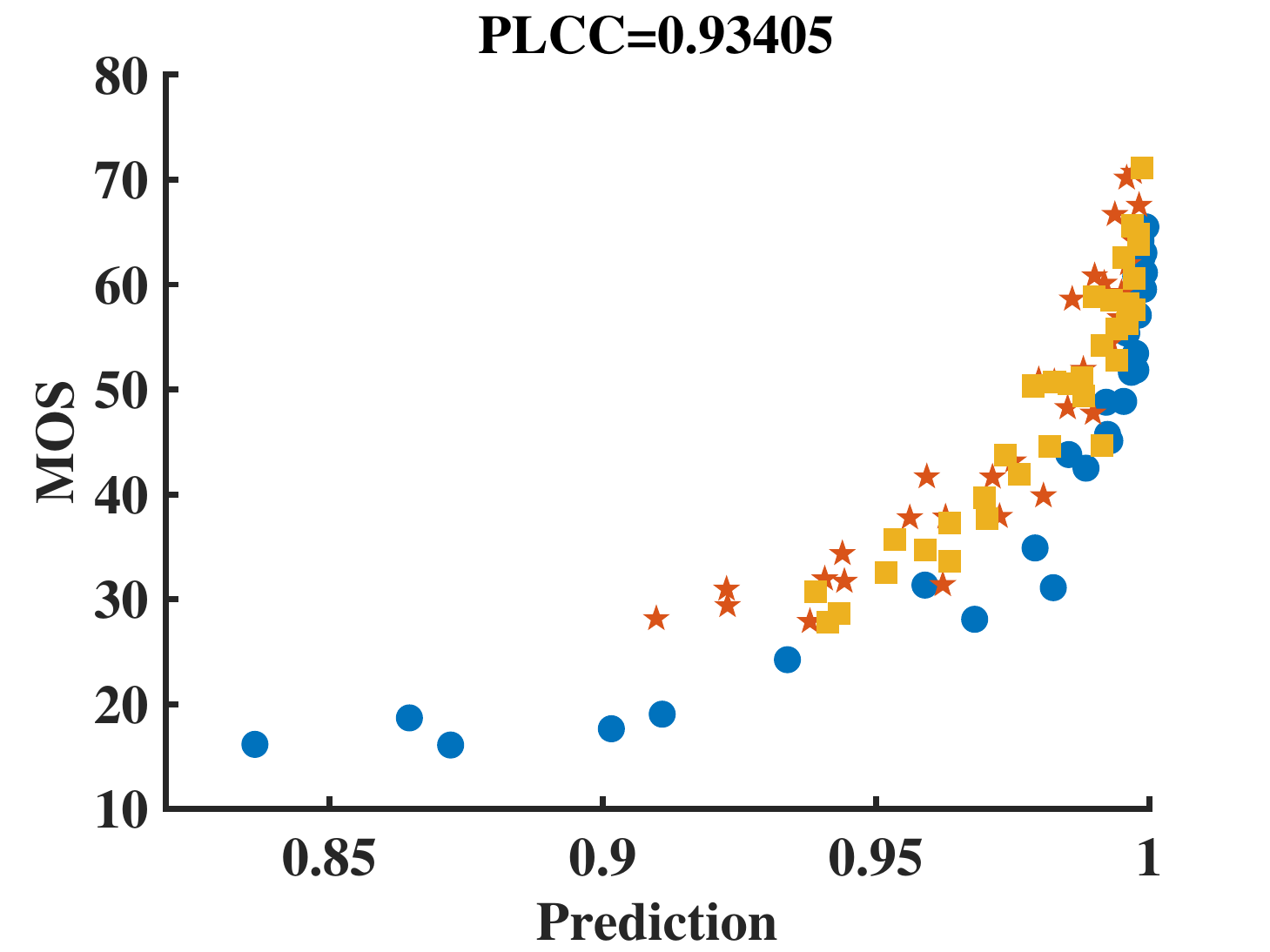}
	}
	\subfigure[DeepQA]{
		\includegraphics[height=2.4cm]{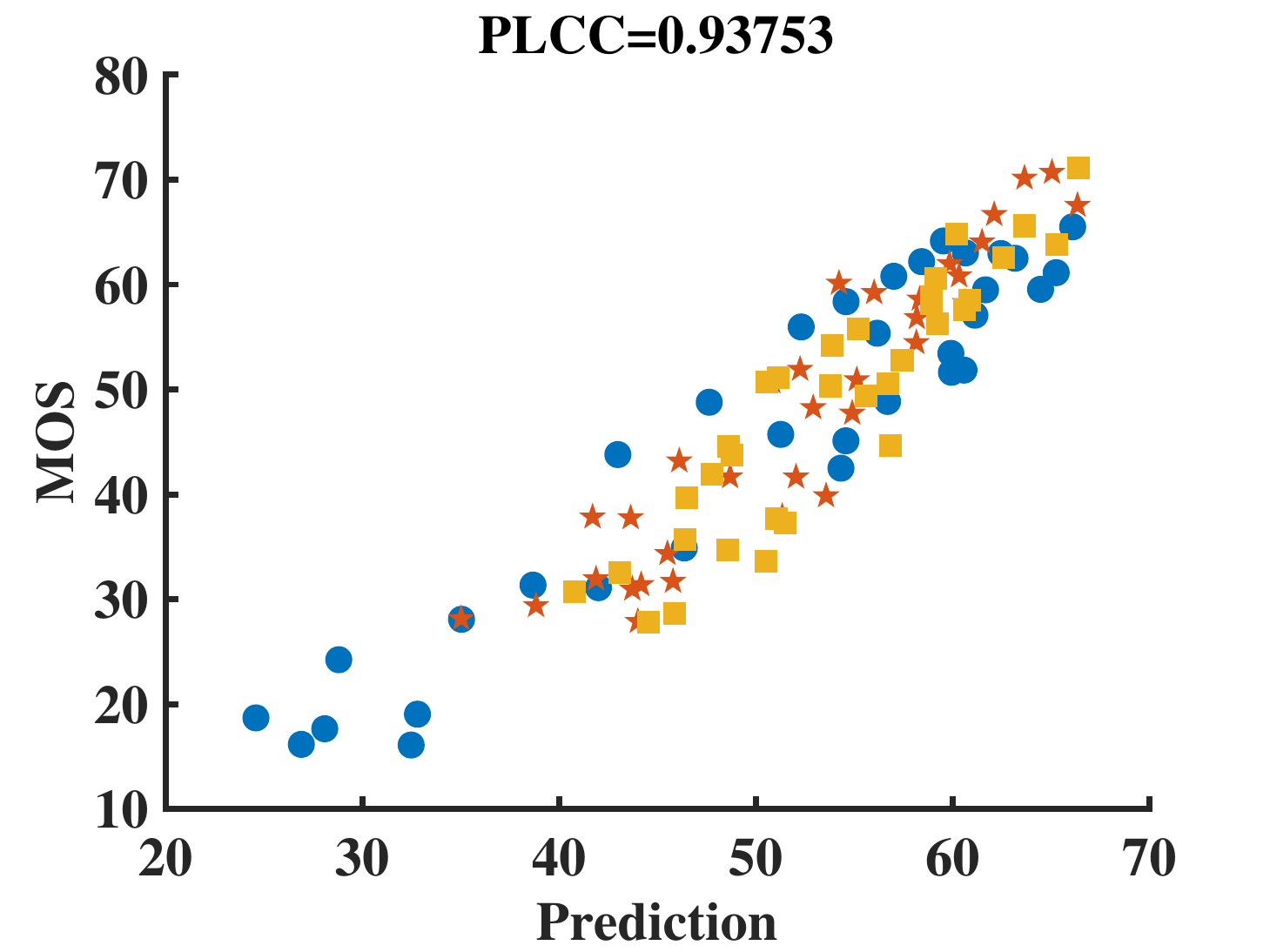}
	}
	\subfigure[BRISQUE]{
		\includegraphics[height=2.4cm]{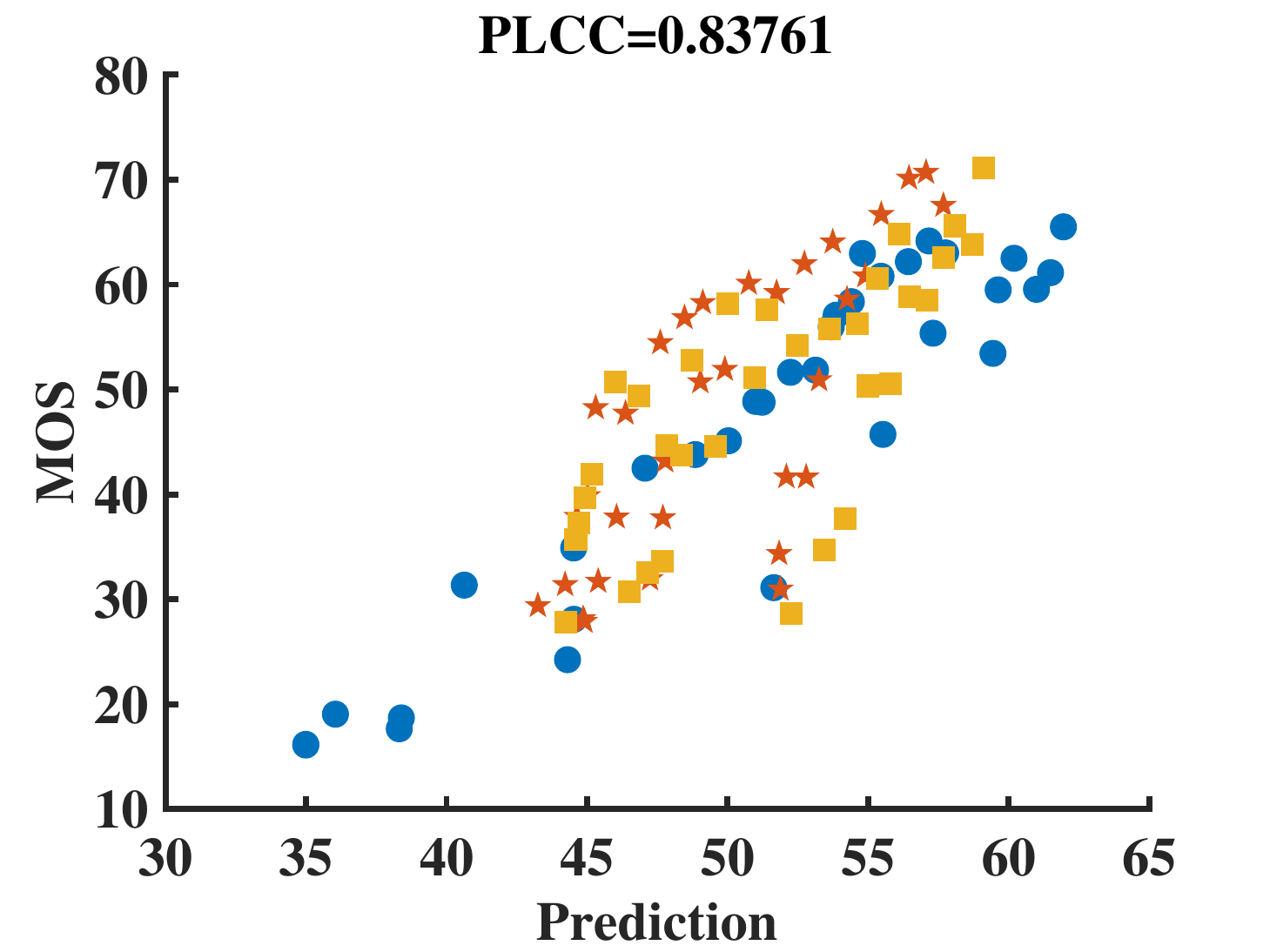}
	}
	\subfigure[BMPRI]{
		\includegraphics[height=2.4cm]{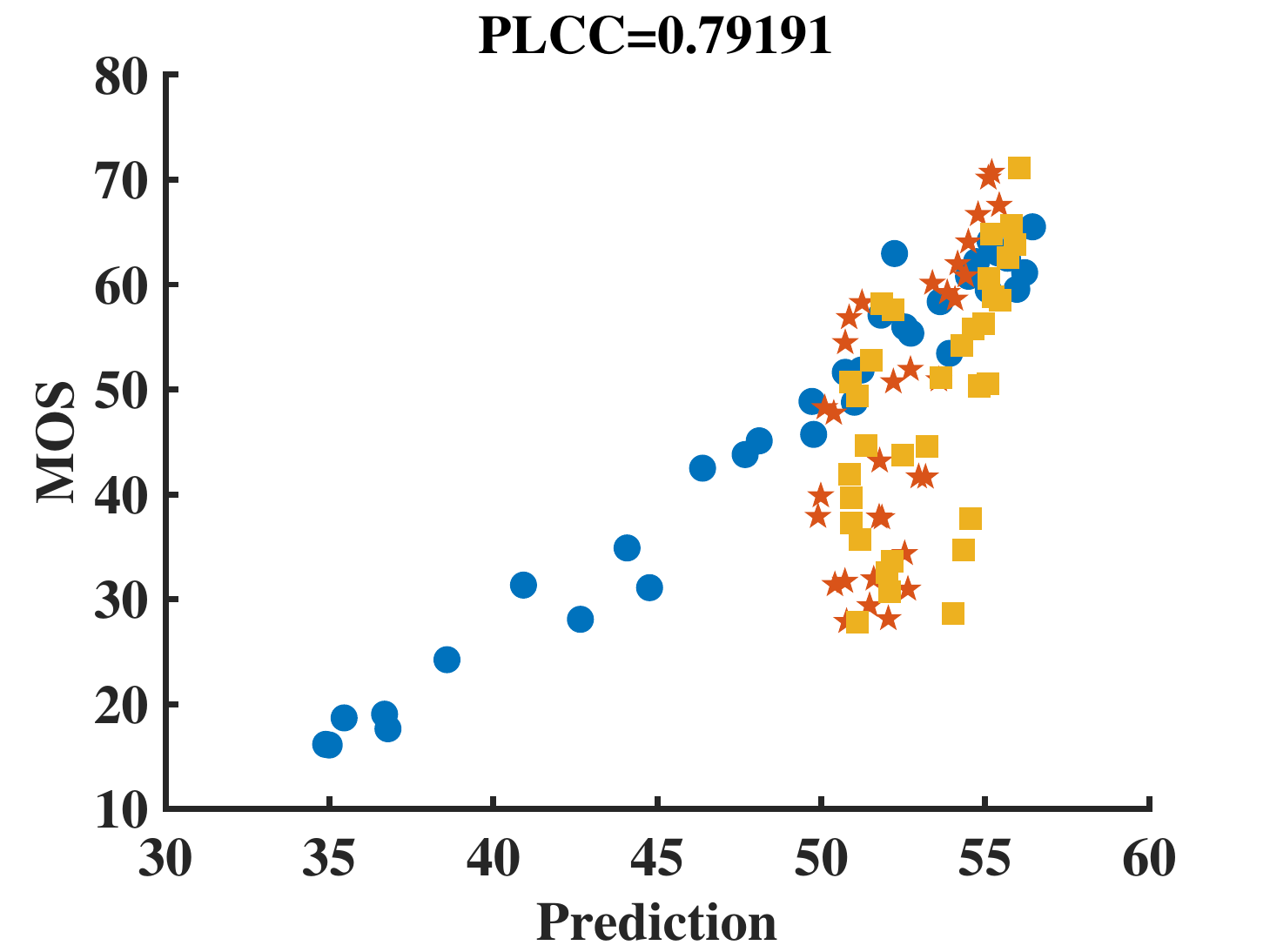}
	}
	\subfigure[DB-CNN]{
		\includegraphics[height=2.4cm]{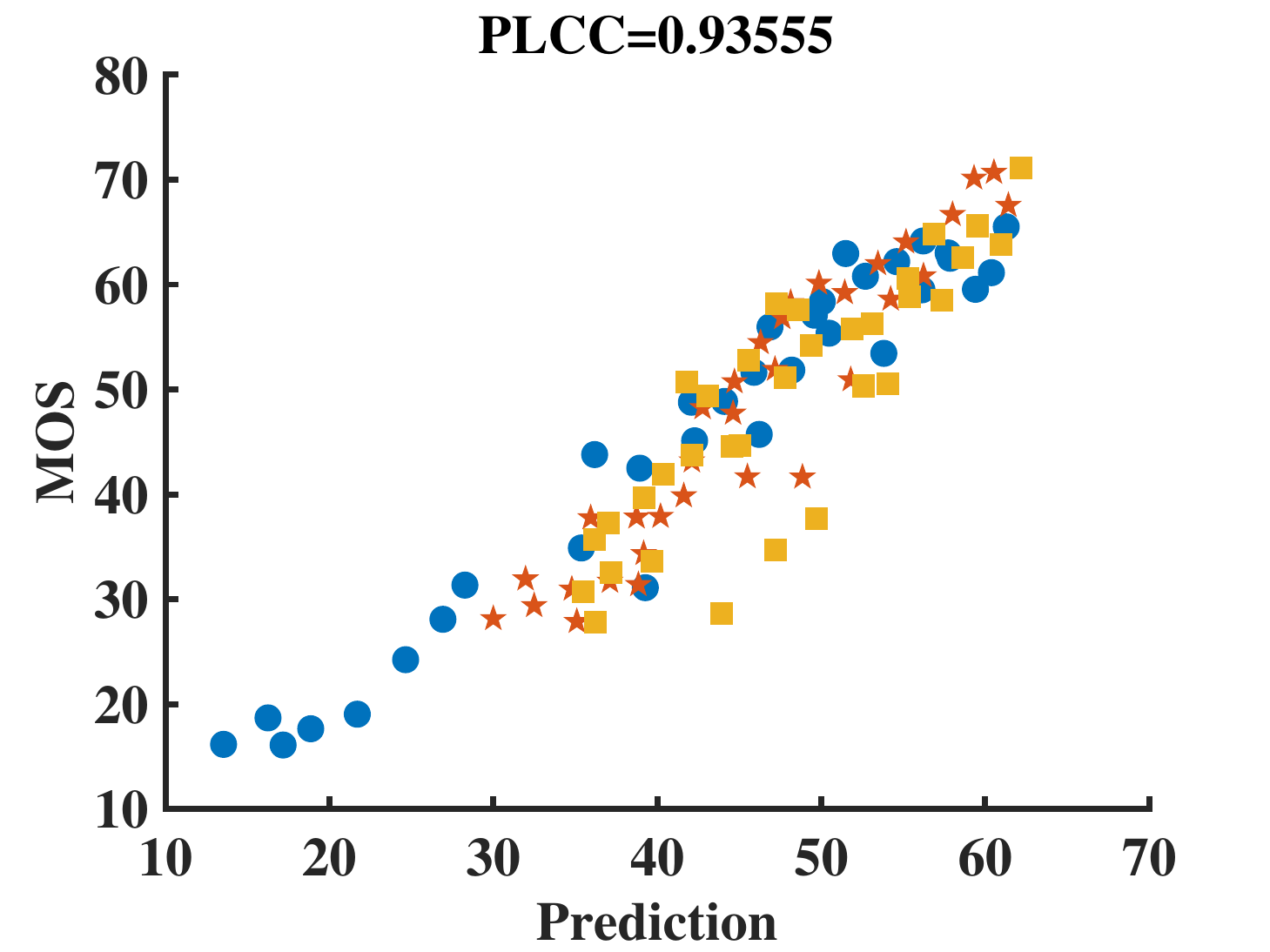}
	}
	\subfigure[MC360IQA]{
		\includegraphics[height=2.4cm]{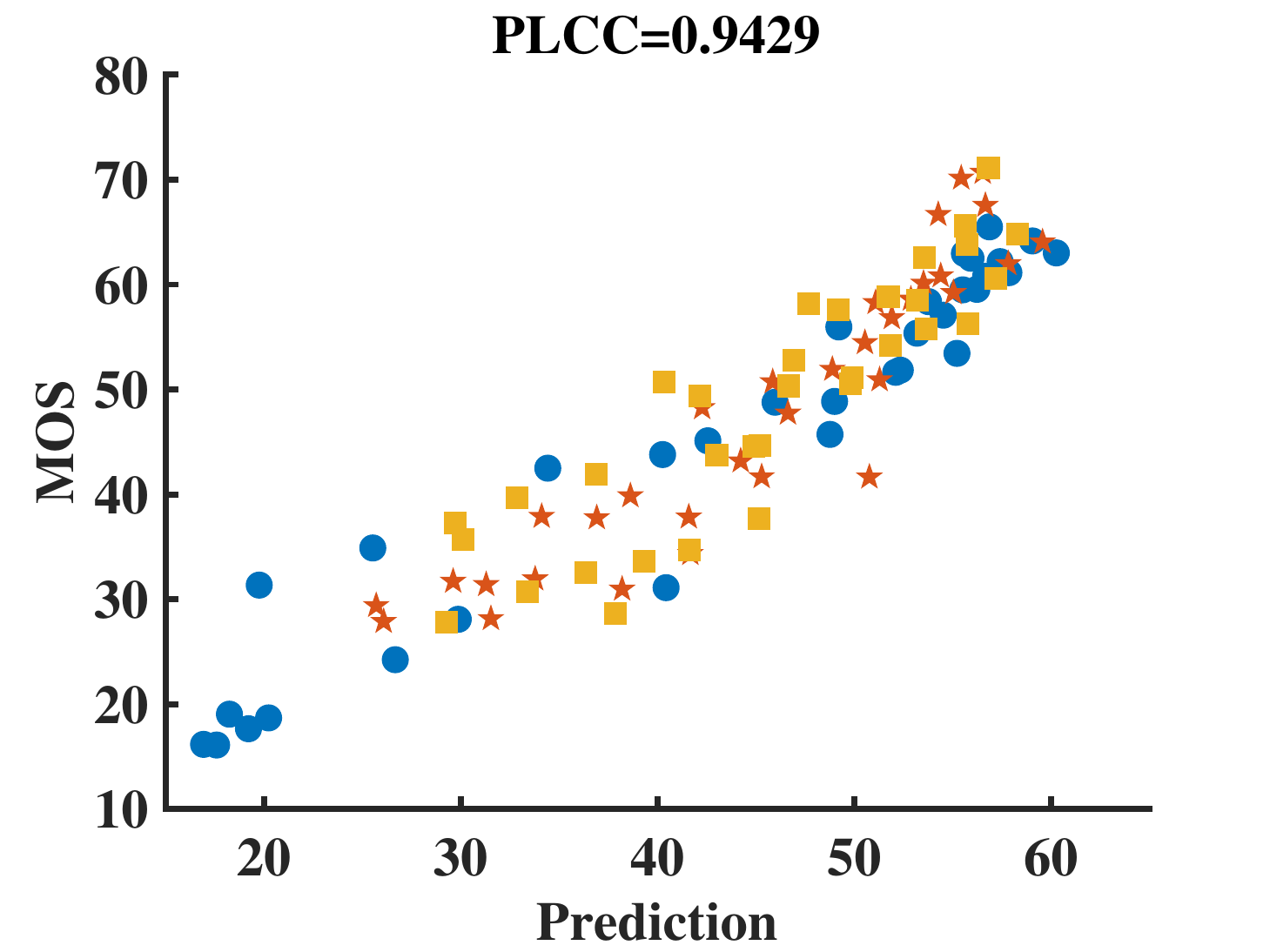}
	}
	\subfigure[VGCN (local)]{
		\includegraphics[height=2.4cm]{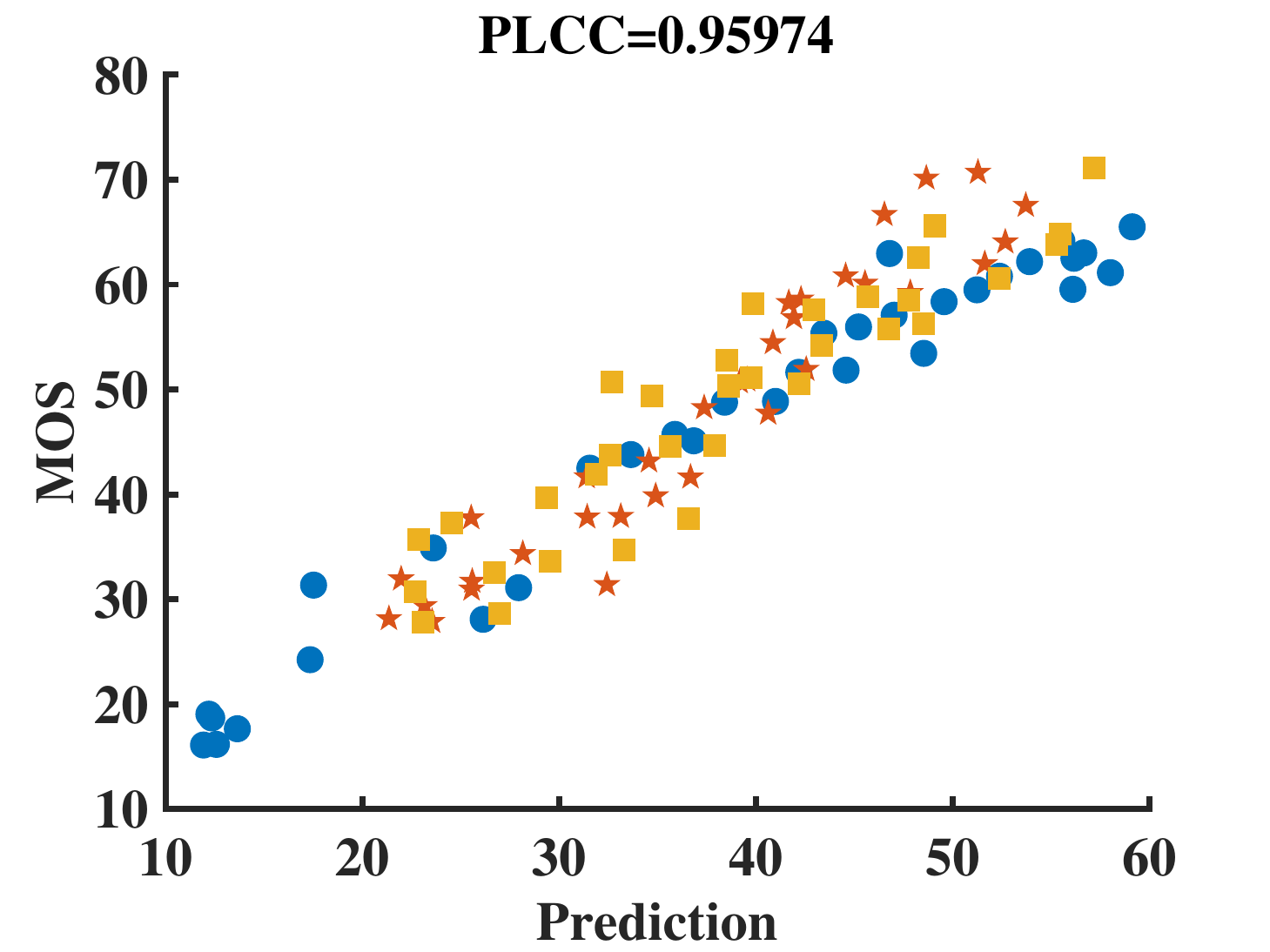}
	}
	\subfigure[AHGCN]{
		\includegraphics[height=2.4cm]{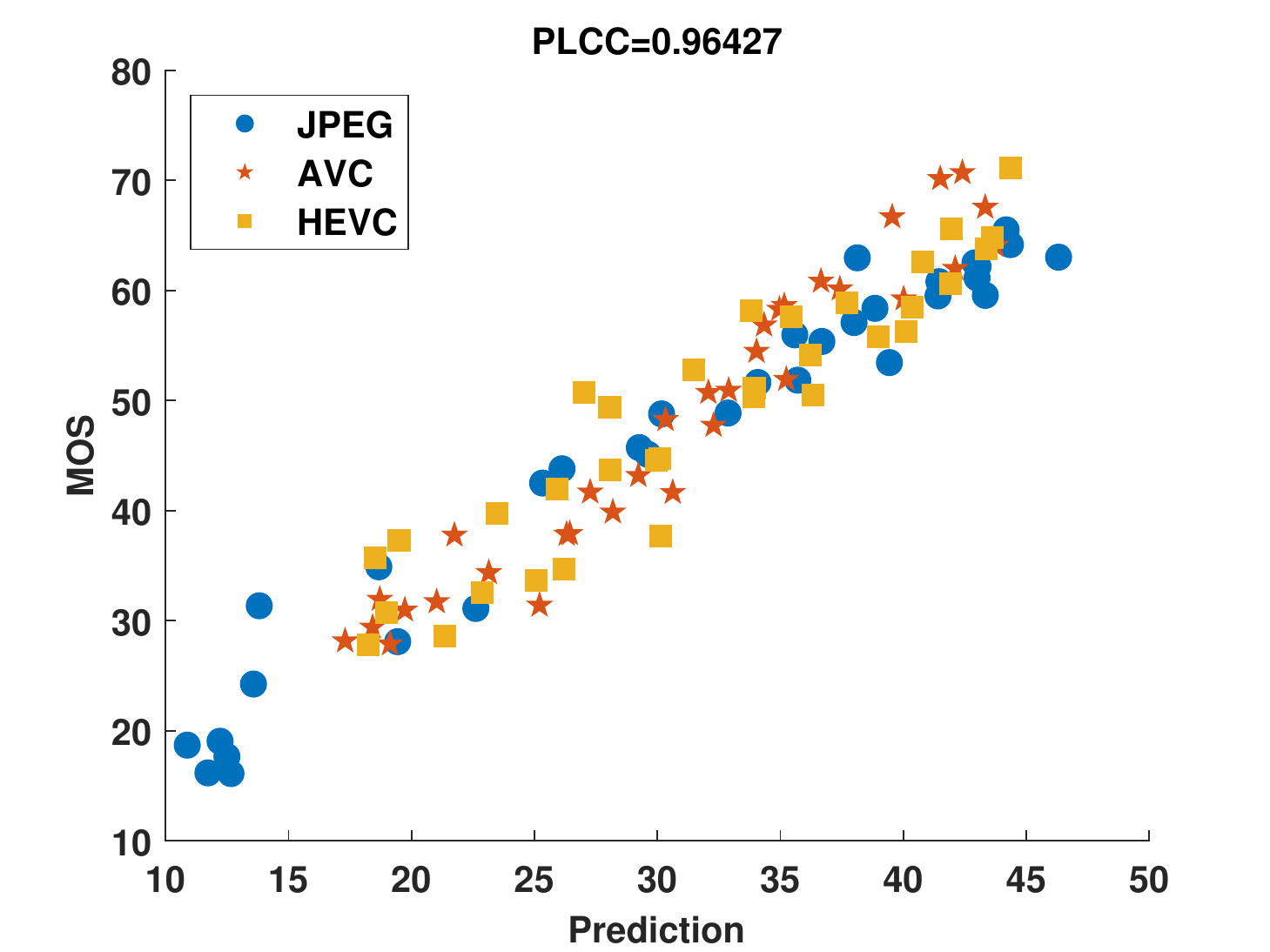}
	}
	\caption{Scatter plots of MOS values versus predictions of IQA metrics on the testing set of CVIQD database. }
	\label{fig:fig5}
\end{figure*}
\begin{figure*}[htbp]
	\renewcommand\arraystretch{1}
	\begin{center}
		\captionsetup{justification=centering}
		\scalebox{0.7}{
			\begin{tabular}{ccccc}
				\includegraphics[width=0.25\linewidth]{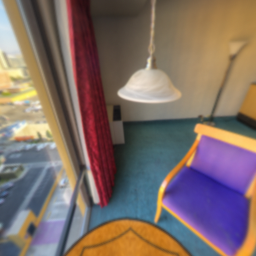} & \includegraphics[width=0.25\linewidth]{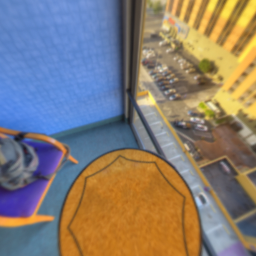} & \includegraphics[width=0.25\linewidth]{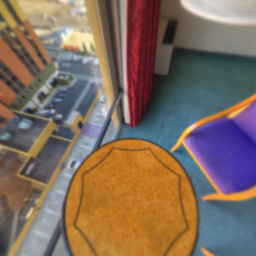}  & \includegraphics[width=0.25\linewidth]{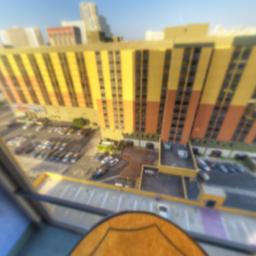}  &
				\includegraphics[width=0.25\linewidth]{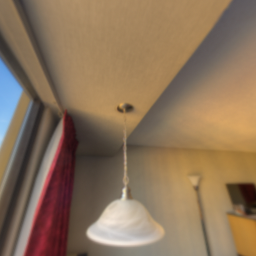}  	\\		
				
				\includegraphics[width=0.25\linewidth]{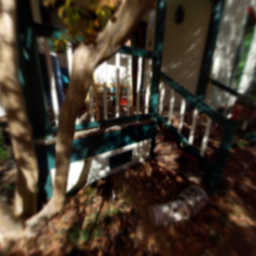} & \includegraphics[width=0.25\linewidth]{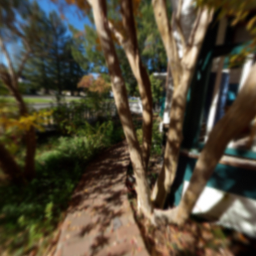} & \includegraphics[width=0.25\linewidth]{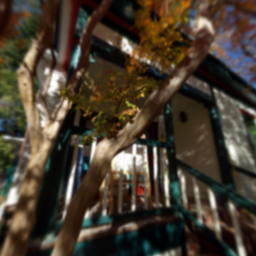}  & \includegraphics[width=0.25\linewidth]{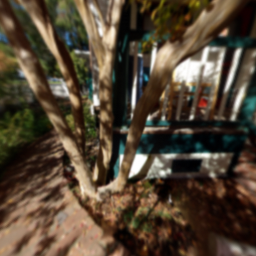}  &
				\includegraphics[width=0.25\linewidth]{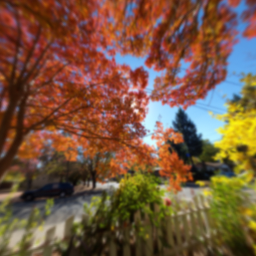}  	\\		
				
				\includegraphics[width=0.25\linewidth]{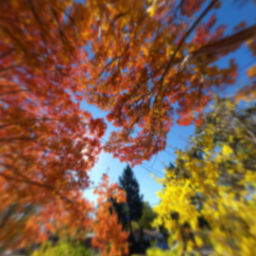} & \includegraphics[width=0.25\linewidth]{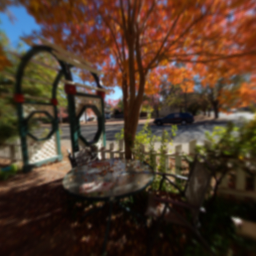} & \includegraphics[width=0.25\linewidth]{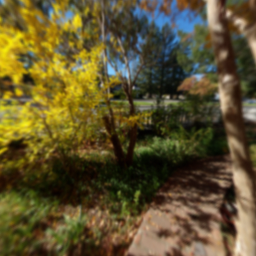}  & \includegraphics[width=0.25\linewidth]{./fig/img239_fov5.png}  &
				\includegraphics[width=0.25\linewidth]{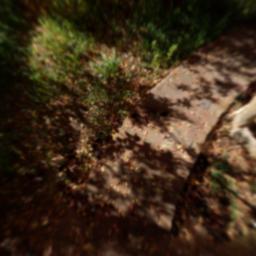}  	\\		
				
				\includegraphics[width=0.25\linewidth]{./fig/img139_fov20.png} & \includegraphics[width=0.25\linewidth]{./fig/img139_fov18.png} & \includegraphics[width=0.25\linewidth]{./fig/img139_fov3.png}  & \includegraphics[width=0.25\linewidth]{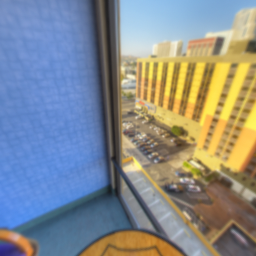}  &
				\includegraphics[width=0.25\linewidth]{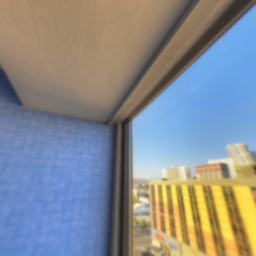}  	\\		
		\end{tabular}}
	\end{center}
	\caption{{Content-based hyperedges on OIQA database. Each row denotes a hyperedge, where the centroid is the first viewport and the remaining viewports are semantic neighborhoods of the centroid.}}
	\label{fig8}
\end{figure*}
\begin{figure}[htbp]
	\centering
	\subfigure[Different vs. Similar (AUC)]{
		\includegraphics[width=7.5cm]{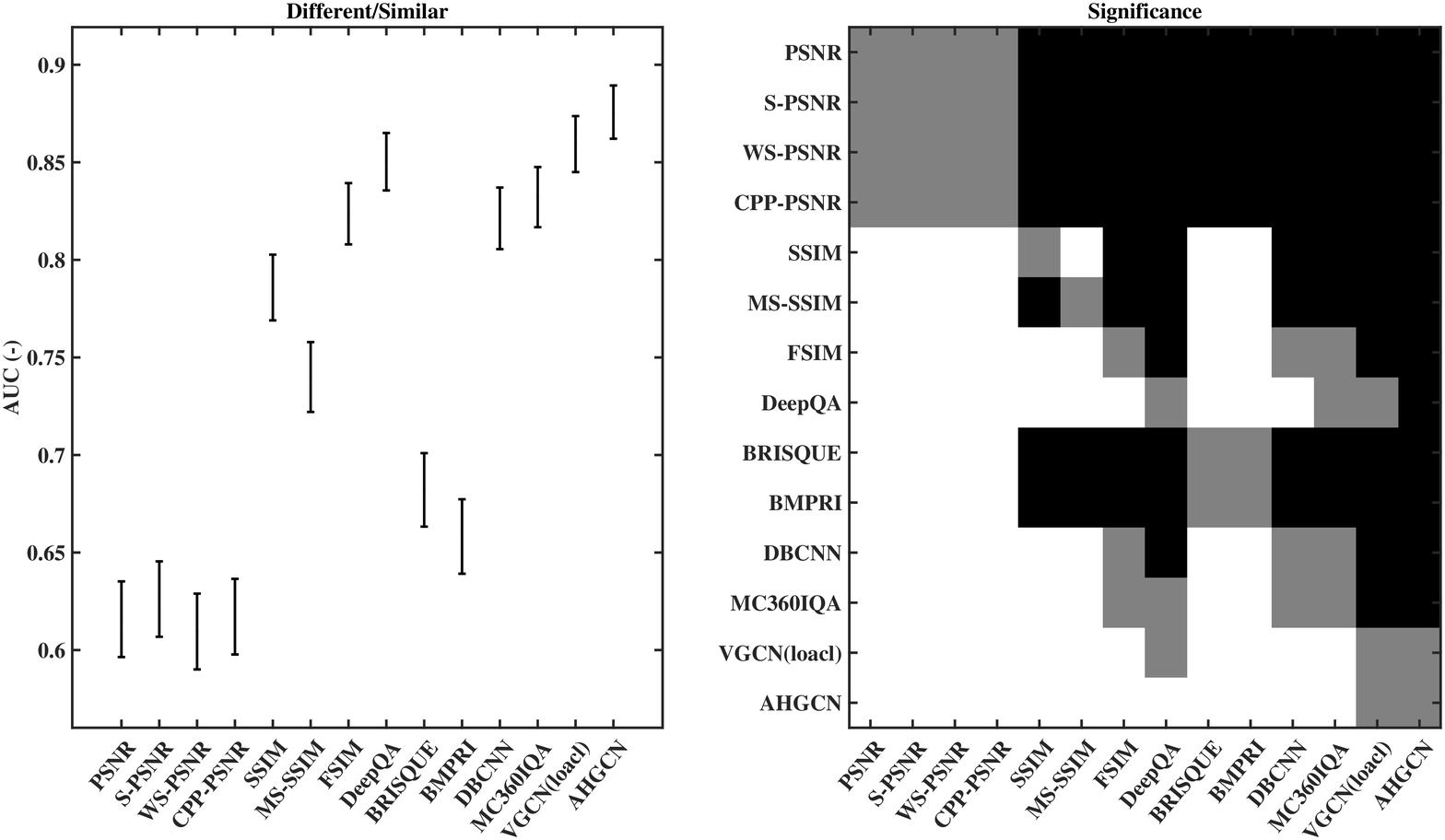}
	}
	\subfigure[Better vs. Worse ($C_0$)]{
		\includegraphics[width=7.5cm]{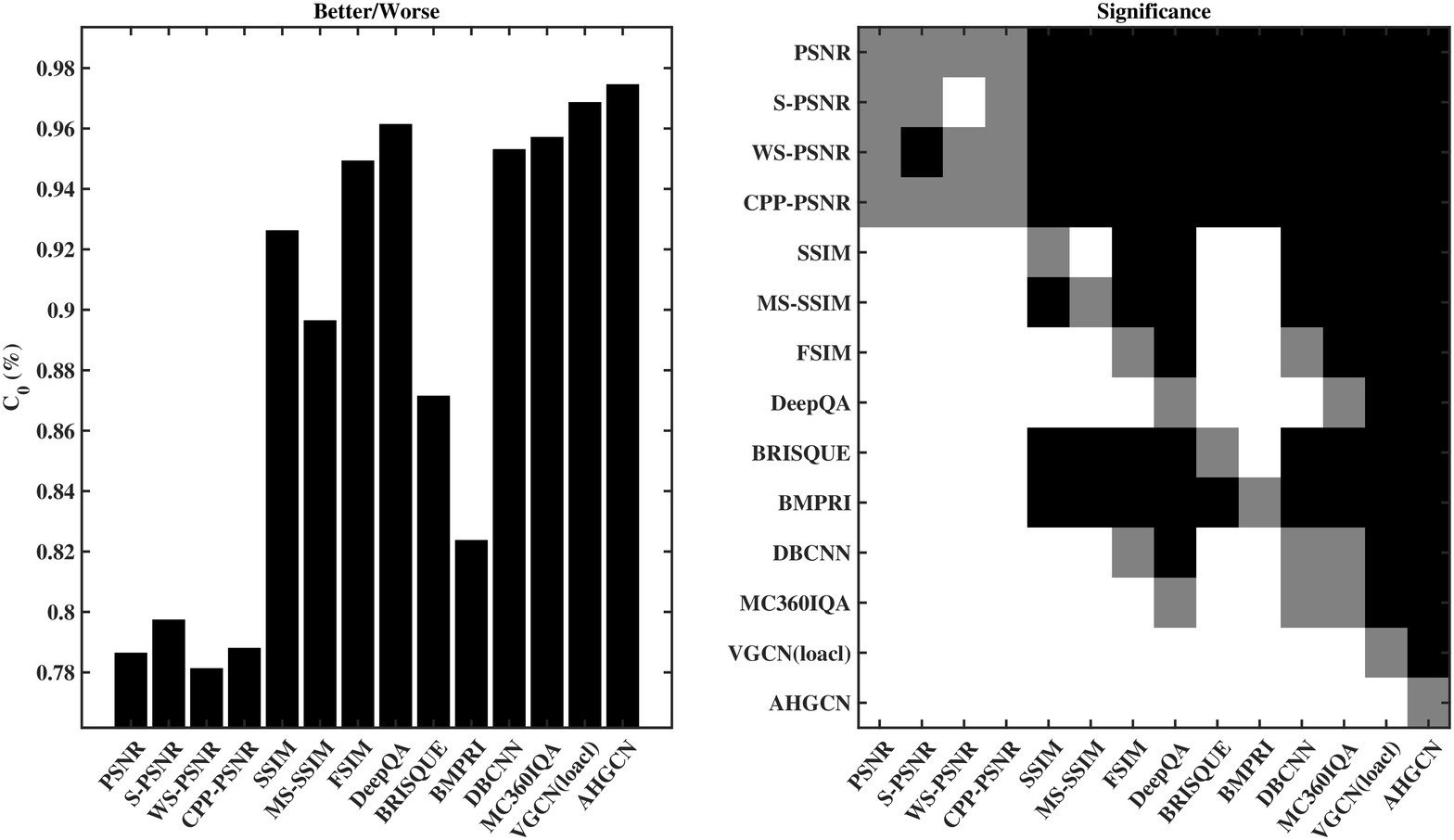}
	}
	\subfigure[Better vs. Worse (AUC)]{
		\includegraphics[width=7.5cm]{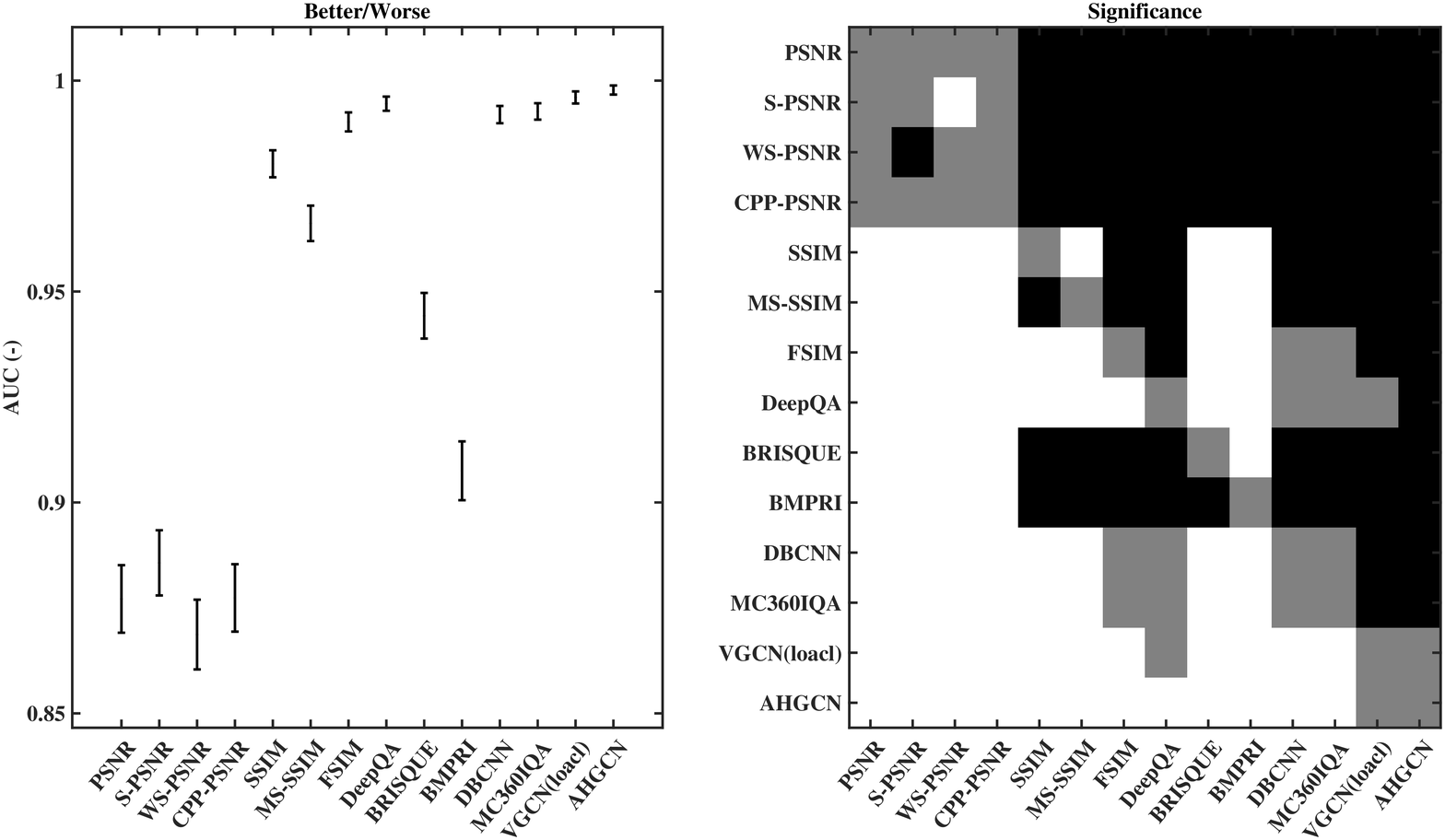}
	}
	\caption{The results of Krasula methodology on the CVIQD database.}
	\label{fig:fig6}
\end{figure}

In this section, we first introduce databases, performance metrics, and baselines. Then, we compare AHGCN with existing competitive IQA metrics on two public datasets. Finally, we conduct an ablation study to verify the effectiveness of each component in AHGCN.

\subsection{Databases}
We evaluate the effectiveness of the proposed method on two public datasets, i.e., OIQA database~\cite{OIQADatabse} and CVIQD database~\cite{CVIQDDatabase}. Following the liteature~\cite{xu2020blind}, we split the database into the training and testing set, and crop 20 viewports for each distorted 360-degree image.

\textbf{OIQA Database:} It contains 320 distorted 360-degree images, which are obtained by applying 4 distortion types with 5 levels to 16 raw 360-degree images. The distortion types are JPEG compression (JPEG), JPEG2000 compression (JP2K), Gaussian blur (BLUR), and Gaussian white noise (WN). Mean opinion score (MOS)  values of distorted 360-degree images range from 1 to 10.

\textbf{CVIQD Database:} It collects 528 compressed 360-degree images generated from 16 lossless source images. It considers three compression distortion types, i.e., JPEG, H.264/AVC, and H.265/HEVC. The MOS values of distorted 360-degree images are normalized and rescaled to the range [0, 100].

\subsection{Evaluation Metrics and Baselines}
Standard measures~\cite{PerformanceMeasure} and the Krasula methodology~\cite{krasula2016accuracy} are used to measure the performance of  IQA metrics.

\textbf{Standard measures:} We choose three standard measures, i.e., Spearmans rank order correlation coefficient (SROCC), Pearsons linear correlation coefficient (PLCC), and root mean squared error (RMSE). PLCC and RMSE reflect the prediction accuracy, while SROCC reflects the prediction monotonicity. A better 360IQA model should have lower RMSE values while higher SROCC and PLCC values. Before calculating the PLCC and RMSE, a five-parameter logistic function~\cite{sheikh2006statistical} is applied on predictions of IQA metrics:
\begin{equation}
\hat{q} = \beta_1\left[ \frac{1}{2} - \frac{1}{1+\text{exp}(\beta_2(q-\beta_3))} \right]+ \beta_4q + \beta_5,
\end{equation}
where $\hat{q}$ is the mapping result of the predicted score $q$. $\beta_1, \beta_2, \beta_3, \beta_4$ and $\beta_5$ are the parameters to be fitted. 

\textbf{Krasula methodology~\cite{krasula2016accuracy}:} The Krasula methodology is used to evaluate the reliability of IQA metrics from three aspects, i.e., the area under the ROC curve of Different vs. Similar (AUC-DS), the area under the ROC curve of  Better vs. Worse categories (AUC-BW), and the percentage of correct classification ($C_0$). Specifically, AUC-BW and AUC-DS represent the capacity of IQA metrics for distinguishing better/worse and different/similar pairs. A better IQA metric should have higher AUC-DS, AUC-BW, and $C_0$. 
\\\\
\noindent\textbf{Baselines:}  We select thirteen representative IQA methods for performance comparison. The competitive approaches include five FR 2DIQA metrics, i.e., PSNR, SSIM~\cite{ssim}, MS-SSIM~\cite{msssim}, FSIM~\cite{fsim}, and DeepQA~\cite{DeepQA}; three learning-based NR 2DIQA metrics, i.e., BRISQUE~\cite{BRISQUE}, BMPRI~\cite{BMPRI}, and DB-CNN~\cite{DBCNN}; three FR 360IQA metrics, i.e., S-PSNR~\cite{SPSNR}, WS-PSNR~\cite{WSPSNR}, and CPP-PSNR~\cite{CPPPSNR}; two viewport-oriented NR 360IQA metrics, i.e., MC360IQA~\cite{mc360iqa2} and VGCN~\cite{xu2020blind}. Compared to MC360IQA, VGCN considers interactions between viewports. It is worth noting that this paper only uses the local branch of the original VGCN for a fair comparison. 

\subsection{Performance Comparison}
Table~\ref{table1} and \ref{table2} present the results of performance comparison. According to the overall performance of each method, we can draw the following conclusions:
\begin{itemize}[noitemsep,topsep=0pt]
	\item PSNR and its variants are inferior to SSIM-based IQA metrics. This is mainly because PSNR-based metrics only reflect the pixel-level distortion, while SSIM-based ones measure the structural distortion related to HVS.
	\item Viewport-oriented NR 360IQA metrics have a clear advantage over FR and NR 2DIQA metrics. This confirms the gap between 2DIQA and 360IQA, and points out the importance of viewport-level information to 360IQA.
	\item VGCN significantly outperforms MC360IQA in terms of PLLC, SROCC, and RMSE. This shows that it is essential to consider interactions between viewports in 360IQA.
	\item The proposed method achieves state-of-the-art performance on both OIQA and CVIQD datasets. Specifically, on the OIQA dataset, compared to VGCN, our approach achieves  0.12,  0.146, and 0.0853 improvements in PLCC, SROCC, and RMSE, respectively. This is mainly owing to two aspects. On the one hand, AHGCN utilizes hierarchical features instead of high-level ones for quality prediction. On the other hand, AHGCN models interactions between viewports through hypergraphs instead of graphs, and consider the semantic relations between viewports. 
\end{itemize}
From the performance of each method in individual distortion type, we have the following observations:
\begin{itemize}
	\item Our proposed AHGCN achieves comparable or superior performance across a broad of distortion types compared with existing competitive IQA metrics.
	\item AHGCN exhibits outstanding power in evaluating compressed 360-degree images, especially for ones encoded by JPEG. 
	\item The development of image compression technology poses more challenges to IQA. As shown in Table~\ref{table2}, the performance of all IQA metrics significantly drops from JPEG to HEVC. One possible reason is that artifacts introduced by new encoders are more indistinguishable compared with blockiness and tonal distortion brought by JPEG.
\end{itemize}

Fig.~\ref{fig:fig4} and ~\ref{fig:fig5} show the scatter plots of ground-truth MOS values versus predictions of IQA models for individual distortion types. We have the following findings:
\begin{itemize}
	\item As shown in Fig.~\ref{fig:fig4}, compared with the other three distortions, predictions of JPEG compression have a lower correlation with subjective scores. This reveals that the quality assessment of compression distortion is more challenging.
	\item According to Fig.~\ref{fig:fig5}, although the difficulty of 360IQA increases with the development of coding technology, our approach achieves a more accurate quality prediction compared with existing NR IQA metrics.
\end{itemize}

Fig.~\ref{fig:fig6} presents the results of Krasula criteria on the CVIQD database.  In the significance plot, black (white) boxes mean that the metric in the row is significantly worse (better) than the metric in the column, and gray boxes indicate that two metrics are evenly matched. As shown in Fig.~\ref{fig:fig6}, our approach achieves the best performance in terms of AUC-DS, AUC-BW, and $C_0$. This reveals that our approach is more reliable than existing IQA metrics. 

\subsection{Visualization on Content-based Hyperedges}
Fig.~\ref{fig8} visualizes four content-based hyperedges on the OIQA database. As we can see,  the centroid viewport and its semantically neighboring viewports typically contain some same or similar objects. As such, evaluating the centroid viewport can resort to its semantic neighborhoods. Unfortunately, the content-based hyperedge may also introduce some disturbance. For example, as shown in the last row in  Fig.~\ref{fig8}, the centroid viewport has a low visual correlation with the other four viewports. It may not be appropriate to add connections between them.

\subsection{Ablation study}
Table~\ref{table3} summarizes the results of the ablation study to prove the effectiveness of each component in AHGCN. According to the table, we draw the following conclusions:
\begin{itemize}
	\item Most of AHGCN's revenue comes from the extracted hierarchical features. For example, on the OIQA dataset, hierarchical features bring 0.0374, 0.0378, and 0.2324 improvements in terms of PLCC, SROCC, and RMSE, respectively.
	\item Compared to AHGCN and the GCN-based 360IQA metric, the baseline using FC for quality prediction achieves much lower performance. This means that it is necessary to consider interactions between viewports in 360IQA.
	\item AHGCN is superior to the GCN-based 360IQA metric on both CVIQD and OIQA datasets. This confirms that hypergraphs perform better in modeling interactions between viewports than graphs. 
	\item For the OIQA dataset, both content-based and location-based hyperedges are beneficial to boost the performance. Moreover, location-based hyperedges bring more gains compared with content-based ones. 
\end{itemize} 
The number of semantic neighborhood $k$ is a key parameter in AHGCN. Table~\ref{table4} presents the performance of AHGCN with different configurations of $k$. From the table, we have the following observations:
\begin{itemize}
	\item The optimal settings of $k$ are $0$ and $5$ on CVIQD and OIQA datasets.
	\item For the OIQA dataset, both an excessively small and large $k$ will weaken the prediction performance.
	\item For the CVIQD dataset, introducing content-based hyperedges will impair the performance of AHGCN. 
\end{itemize} 
\begin{table}[htbp]
	\renewcommand\arraystretch{1.5}
	\begin{center}
		\captionsetup{justification=centering}
		\caption{\textsc{Ablation study on each proposed component in AHGCN. AHGCN only uses location-based hyperedges on the CVIQD dataset.}}
		\label{table3}
		\scalebox{0.7}{
			\begin{tabular}{@{}l|ccc|ccc@{}}
				\toprule
				& \multicolumn{3}{c|}{CVIQD}          & \multicolumn{3}{c}{OIQA}           \\ \midrule
				& PLCC            & SROCC           & RMSE            & PLCC            & SROCC           & RMSE            \\ \midrule
				AHGCN   & \textbf{0.9643} & \textbf{0.9623} & \textbf{3.6990}       & \textbf{0.9649} & \textbf{0.9590} & \textbf{0.5487}           \\\midrule
				---Hierarchical features  & 0.9595          & 0.9552          & 3.9323        & 0.9275          & 0.9212        & 0.7811          \\
				---Content-based hyperedges     & \textbf{0.9643} & \textbf{0.9623} & \textbf{3.6990}          &0.9621         & 0.9554          &   0.5699       \\
				---Location-based hyperedges    & -        & -         & -       &0.9550         & 0.9482         & 0.6196          \\ \midrule
				Hierarchical features  + FC     & 0.9542          & 0.9465          & 4.1771         & 0.9379          & 0.9207        & 0.7257          \\		
				Hierarchical features  + GCN   & 0.9609          & 0.9564        & 3.8657         & 0.9619          & 0.9553       & 0.5711          \\
				\bottomrule
		\end{tabular}}
	\end{center}
\end{table}

\begin{table}[htbp]
	
	\renewcommand\arraystretch{1.5}
	\begin{center}
		\captionsetup{justification=centering}
		\caption{\textsc{Ablation study on the number of semantic neighborhood $k$.}}
		\label{table4}
		\scalebox{0.6}{
			\begin{tabular}{@{}c|ccc|ccc@{}}
				\toprule
				& \multicolumn{3}{c|}{CVIQD}          & \multicolumn{3}{c}{OIQA}           \\ \midrule
				The number of semantic neighborhood $k$ & PLCC            & SROCC           & RMSE            & PLCC            & SROCC           & RMSE            \\ \midrule
				0     & \textbf{0.9643} & \textbf{0.9623} & \textbf{3.6990}          & 0.9621         & 0.9554          &   0.5699    \\
				5     & 0.9593        & 0.9540          & 3.9445       &  \textbf{0.9649} & \textbf{0.9590} & \textbf{0.5487}        \\
				10     & 0.9538        & 0.9514         &4.1929      & 0.9624     & 0.9581          & 0.5673         \\				
				15   & {0.9566} & {0.9506 }         & 4.0695    & 0.9624    & 0.9574          & 0.5673         \\
				20   & 0.9559        & 0.9535         & 4.1016       &0.9617         & 0.9520        & 0.5726         \\
				\bottomrule
		\end{tabular}}
	\end{center}
\end{table}


\section{Conclusion}
In this paper, we propose an adaptive hypergraph convolutional neural network for NR 360IQA, dubbed as AHGCN. It designs a multi-level viewport descriptor to extract hierarchical representations from viewports, and models interactions between viewports through hypergraphs instead of graphs. In the hypergraph construction, it considers both the locations and content features of viewports. Experimental results demonstrate that the proposed AHGCN achieves state-of-the-art performance and shows an impressive generalization capability across a broad of distortion types. In the future, we first plan to explore an adaptive kNN algorithm for content-based hyperedge construction. Then, we will focus on extending hypergraphs to 360-degree video quality assessment (360VQA) for capturing complex spatial-temporal interactions between viewports. Finally, we will try to apply hypergraphs into other IQA tasks, such as stereoscopic IQA~\cite{zhou2021hierarchical,xu2020binocular,chen2020stereoscopic,zhou2019dual}.


%

%

\section*{Acknowledgment}
This work was supported in part by NSFC under Grant U1908209,  61632001 and the National Key Research and Development Program of China 2018AAA0101400.

\bibliographystyle{IEEEtran}
\bibliography{references}

\end{document}